\documentclass[aps,twocolumn,nofootinbib]{revtex4}
\usepackage{amsmath,epsfig}
\usepackage{color}
\pdfoutput=1

\def\bsp#1\esp{\begin{split}#1\end{split}}

\begin{document}

\vspace*{-2.5cm}
\vspace*{0.5cm}
\begin{flushright}
IFT-UAM/CSIC-19-129 \\
CERN-TH-2019-170
\end{flushright}
\vspace{0.cm}

\title{New physics with boosted single top production at the LHC and future colliders}

\author{J.A.~Aguilar--Saavedra$^{(a,b)}$, M.L. Mangano$^{(c)}$}

\affiliation{{\phantom.}\\
$^{(a)}$ \mbox{Instituto de F\'isica Te\'orica UAM-CSIC, Campus de Cantoblanco, E-28049 Madrid, Spain}\\
$^{(b)}$ \mbox{Universidad de Granada, E-18071 Granada, Spain (on leave)} \\
$^{(c)}$ \mbox{CERN, TH Department, CH-1211 Geneva 23, Switzerland}
}

\begin{abstract}
We address the potential of measurements with boosted single-top final states at the high-luminosity LHC (HL-LHC) and possible future hadron colliders: the high-energy LHC (HE-LHC), and the future circular collider (FCC). As new physics examples to assess the potential, we consider the search for $tbW$ anomalous couplings and for a weakly-coupled $W'$ boson. The FCC would improve by a factor of two the sensitivity to anomalous couplings of the HL-LHC. For $W'$ bosons, the FCC is sensitive to $W'$ couplings $2-5$ times smaller than the HL-LHC in the mass range 2-4 TeV, and to masses up to 30~TeV in the case of Standard Model-like couplings.  
 \end{abstract}

\maketitle

\section{Introduction}

Future proton-proton colliders with higher energy and luminosity than the Large Hadron Collider (LHC) will push the energy frontier to the multi-TeV scale. In addition, precision measurements will be possible near the TeV scale, which are not currently feasible at the LHC because of insufficient statistics. These precision measurements have high relevance and are complementary to the searches performed at the kinematical end of the spectrum. As is well known, precision measurements provide indirect tests of the presence of new physics, too heavy to be directly detected. But precision measurements can also probe new weakly-interacting resonances at the TeV scale, with cross sections too small to be detected at the LHC.

Both physics cases will be addressed in this paper. We study the potential of single-top plus jet final states to probe anomalous $tbW$ couplings and new $W'$ bosons decaying into $tb$, at the high-luminosity (HL-LHC~\cite{Azzi:2019yne}) and a possible high-energy (HE-LHC) upgrade of the LHC with centre-of-mass (CM) energy $\sqrt{s}=27$\;TeV~\cite{Abada:2019ono}, as well as at a potential future circular collider (FCC), colliding protons at $\sqrt{s}=100$\;TeV~\cite{Abada:2019lih,Benedikt:2018csr}. Because the decay of boosted top quarks yields a single fat jet in the detector, the measurement of Standard Model (SM) single-top production, either in the $t$-channel ($tj$) or $s$-channel ($tb$) processes, is already a daunting task, requiring the reduction of huge backgrounds: light jet pairs, $b \bar b$ and $t \bar t$. In Section~\ref{sec:2} we present our analysis for the HL-LHC with 14 TeV, considering SM single top production as signal and the rest of processes as background. The first step of the analysis is thus to reduce these backgrounds in order to maximise the significance of SM single top signals. In Sections~\ref{sec:3} and \ref{sec:3b} we do the same for HE-LHC with 27 TeV and the FCC with 100 TeV, respectively.

The signals we are interested in, ultimately, produce deviations with respect to the SM prediction for single top production. The strategies to detect non-resonant anomalous top interactions or a resonant $W'$ boson are different. Relying on the analyses presented in Sections~\ref{sec:2}--\ref{sec:3b}, we present in Section~\ref{sec:4} the limits on a particular type of $tbW$ anomalous coupling, to which top decay angular distributions have little sensitivity~\cite{AguilarSaavedra:2006fy,AguilarSaavedra:2008gt} (see also~\cite{Godbole:2018wfy}). In Section~\ref{sec:5} we discuss the sensitivity to weakly-coupled TeV-scale $W'$ bosons. We complete our review with the study in Section~\ref{sec:a} of the mass reach for $W'$ bosons with $O(1)$ coupling at the FCC. The discussion of our results is given in Section~\ref{sec:6}.

\section{Analysis at 14 TeV}
\label{sec:2}
The various processes involved are generated using {\scshape MadGraph5}~\cite{Alwall:2014hca}, followed by hadronisation and parton showering with {\scshape Pythia~8}~\cite{Sjostrand:2007gs} and detector simulation using {\scshape Delphes 3.4}~\cite{deFavereau:2013fsa}. The detector card corresponds to the basic performance of the upgraded ATLAS and CMS detectors~\cite{card}, modified to remove the isolation criteria for electrons and muons so as to include the non-isolated charged leptons in our analysis. Signal processes are $t$-channel single top quark (and anti-quark) production in the five-flavour scheme (labelled as `$tj$'), and $s$-channel production (labelled as `$tb$'). The main backgrounds are $t \bar t$, $b \bar b$ and light dijet production, labelled as $jj$.
The events are generated by dividing the phase space in narrow 100 GeV slices of transverse momentum $p_T$, starting at $p_T \in [400,500]$ GeV, and with the last bin at $p_T \geq 1300$ GeV. In each slice, $2\times 10^4$ events are generated for $tb$, $10^5$ events for each of the $tj$, $t \bar t$ and $b \bar b$ processes and $3 \times 10^5$ events for $jj$. All the decay modes of the top quarks are included. The CM energy is set to 14 TeV, with the expected luminosity of $\mathcal{L} = 3~\text{ab}^{-1}$.

For this analysis we use two main collections of jets, fat jets of radius $R = 0.8$ and narrow jets with $R=0.4$, reconstructed using the anti-$k_T$ algorithm~\cite{Cacciari:2008gp}.\footnote{These collections are independently obtained by clustering particles into jets of maximum radii $R=0.8$, $R=0.4$.} Fat jets are trimmed~\cite{Krohn:2009th} using the parameters $R=0.2$, $f_\text{cut} = 0.05$ to eliminate the contamination from initial state radiation, underlying event and multiple hadron scattering. 
For narrow jets we use Soft Drop~\cite{Larkoski:2014wba} with parameters $z_\text{cut} = 0.1$, $\beta = 0$.
Jet reconstruction and grooming is performed with {\scshape FastJet}~\cite{Cacciari:2011ma}. We do not include pile-up in the simulation. For the jet mass, the effect was previously shown to be small~\cite{Aguilar-Saavedra:2017vka} by comparing the mass for top jets, with and without pile-up. For the jet substructure analysis we use the ungroomed jets. We assume that the pile-up contamination can conveniently be removed by using tools such as {\scshape Puppi}~\cite{Bertolini:2014bba}, widely used by the CMS Collaboration,  {\scshape Softkiller}~\cite{Cacciari:2014gra} or constituent level subtraction~\cite{Berta:2014eza}, previous to the analysis of jet substructure of the ungroomed jets.

We select the $R=0.8$ jet with highest trimmed mass among the two ones with highest transverse momentum $p_T$, and label it as `top' jet $J$. This jet is required to be very energetic, with $p_{TJ} \geq 500$ GeV. A light jet $j$ is selected among the $R=0.4$ jets as the one with highest $p_T$ that has azimuthal separation $\Delta \phi_{Jj} \geq 2.5$ from the top jet in the plane transverse to the beam axis. It is then required that both jets have pseudo-rapidity $|\eta| \leq 2.5$. This latter requirement is almost fully efficient for single top production. Even for the case of the $t$-channel process, at high $p_{TJ} \geq 500$ GeV the pseudo-rapidity distribution is rather central and therefore the pseudo-rapidity cut has little effect on the cross section, keeping 98\% of the events. In order to reduce the huge dijet background, which amounts to $520$ pb after this selection, we require the presence of a charged lepton within a cone $\Delta R \leq 0.5$ of the top jet. If there is more than one lepton in the event, the leading one is selected. In top quark decays (either in the single top signals or in the $t \bar t$ background) energetic leptons result from the leptonic decay of the $W$ boson, while less energetic leptons can also result from $b$ quark decays.
Requiring the presence of a charged lepton reduces the dijet background to $33$ pb, and the efficiency for the $t$-channel and $s$-channel signals are $0.24$ and $0.25$, respectively.
We will hereafter refer to this set of pre-selection cuts as `topology cuts'.

\begin{figure}[htb]
\begin{center}
\includegraphics[width=8cm]{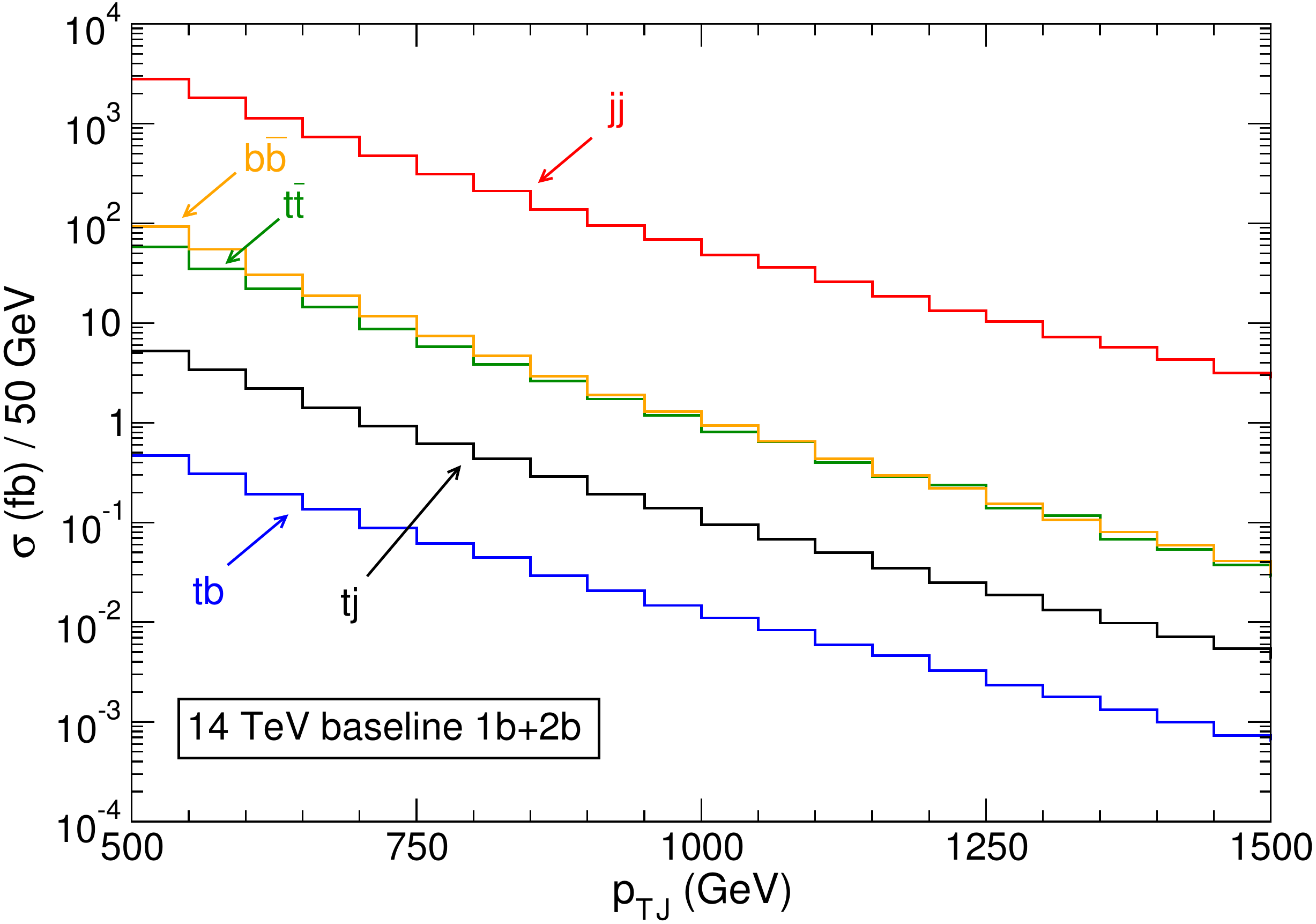} \\
\caption{Signal and background cross sections as a function of the transverse momentum of the `top' jet $p_{T J}$ for a CM energy of 14 TeV, with the baseline event selection.}
\label{fig:dist0}
\end{center}
\end{figure}

The dijet background can be further reduced by $b$-tagging the top jet. A third collection of `track jets' of radius $R=0.2$, reconstructed using only tracks, is used, and the top jet is considered as $b$-tagged if a $b$-tagged track jet (using the 75\% efficiency working point) within $\Delta R = 0.2$ of its centre is found. This procedure has been previously used, for example, for the tagging of boosted Higgs bosons from the decay of a heavy resonance~\cite{Aaboud:2017cxo}. After $b$-tagging, the dijet background is reduced to $8.2$ pb, and the efficiency for the $t$-channel signal is $0.7$. With this `baseline' event selection, we use $b$-tagging on the light $R=0.4$ jet, again using the 75\% efficiency working point. The sample is then split depending on whether this jet is $b$-tagged or not. The latter, labelled as `$1b$', has a larger fraction of $t$-channel single top production, whereas in the former, labelled as `$2b$', this signal is suppressed by the second $b$ tag, and the contribution from $s$-channel production is larger. The signal and background cross sections in each sample are collected in the first two columns of Table~\ref{tab:sig}. The dependence of the cross sections on the transverse momentum of the top jet is shown in Fig.~\ref{fig:dist0}.

\begin{table}[htb]
\begin{center}
\begin{tabular}{ccccccccc}
& \multicolumn{2}{c}{baseline} & \multicolumn{2}{c}{(i) only}  & \multicolumn{2}{c}{(ii) only} & \multicolumn{2}{c}{(i) $+$ (ii)} \\
& $1b$ & $2b$ & $1b$ & $2b$ & $1b$ & $2b$ & $1b$ & $2b$ \\
$tj$ & 14.6 & 0.561
       & 9.63 & 0.340
       & 11.0 & 0.277
       & 7.25 & 0.185
 \\
$tb$ & $0.525$ & $0.875$ 
        & 0.288 & 0.522 
        & 0.279 & 0.532 
        & 0.170 & 0.335 
\\
$t \bar t$ & 95.3 & 60.5
               & 67.0 & 42.0
               & 14.9 & 8.32
               & 10.3 & 5.66
\\
$b \bar b$ & 81.5 & 148
                 & 12.0 & 22.9
                 & 52.6 & 103
                 & 7.62 & 15.7
\\
$jj$ & 7680 & 264
       & 2050 & 90.2
       & 4940 & 143
       & 1210 & 44.3
\\
\end{tabular}
\caption{Cross sections (in fb) for signals and backgrounds at different stages of event selection, for a CM energy of 14 TeV.}
\label{tab:sig}
\end{center}
\end{table}

\begin{figure}[t]
\begin{center}
\begin{tabular}{c}
\includegraphics[width=8cm]{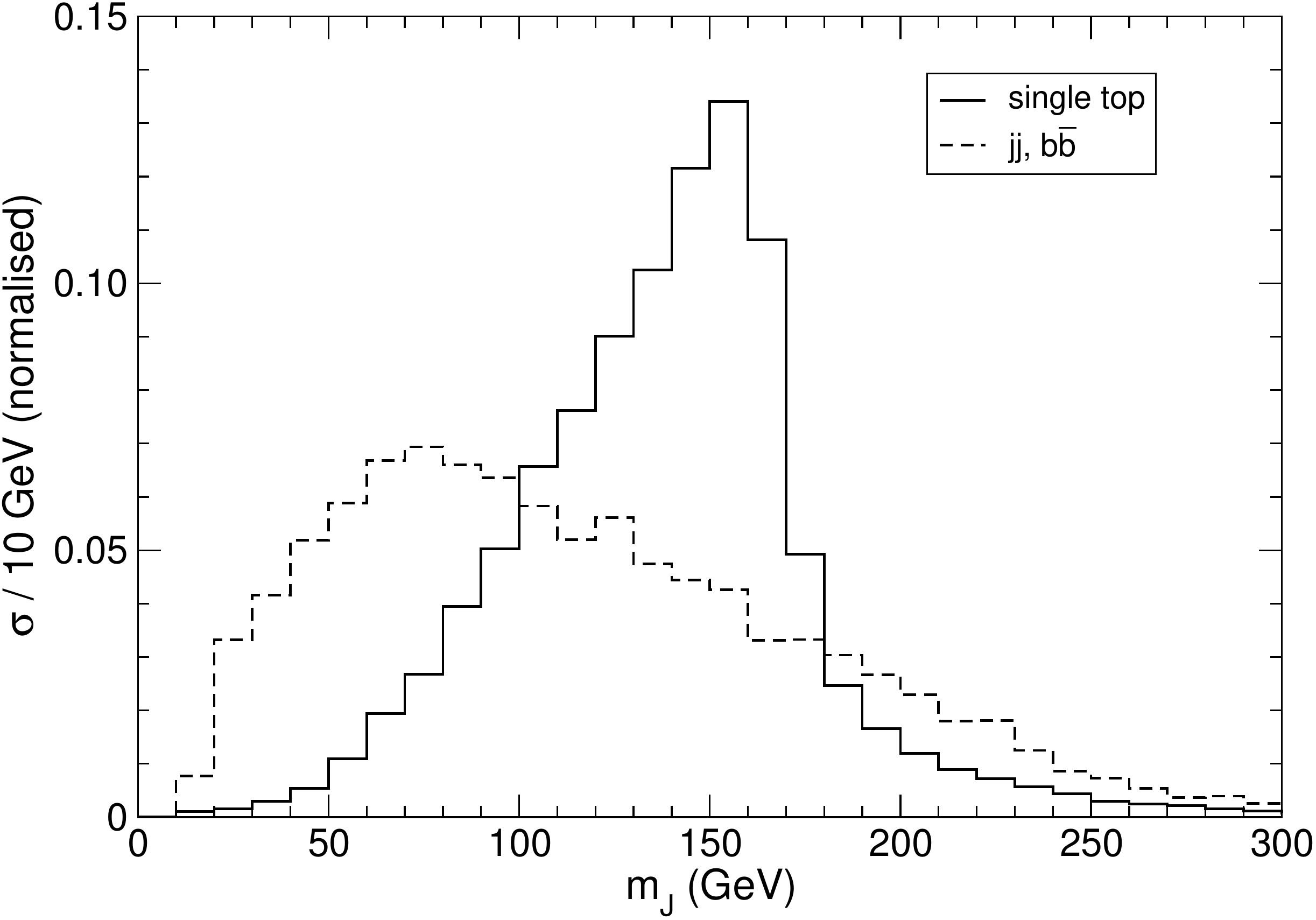} \\
\includegraphics[width=8cm]{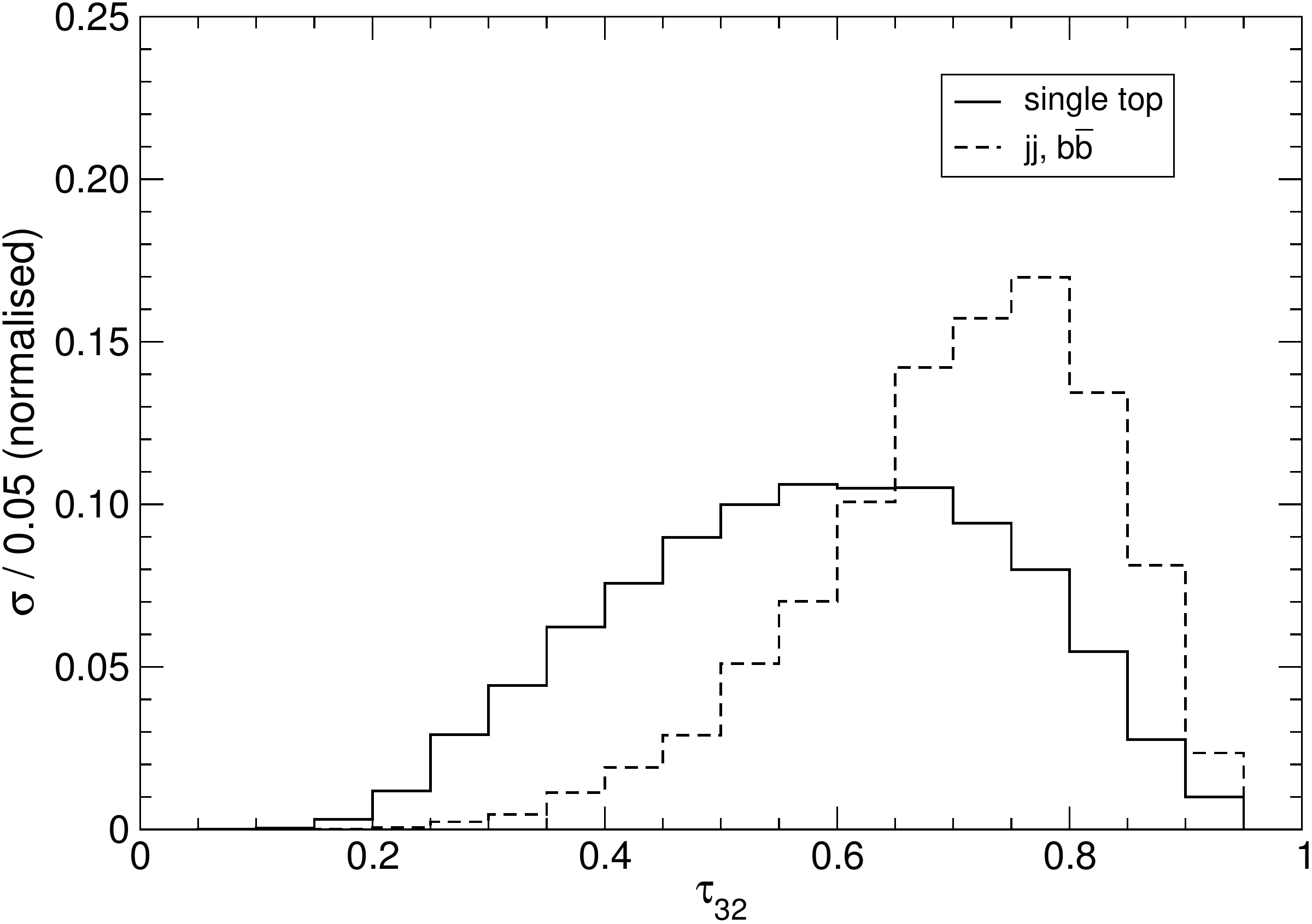}
\end{tabular}
\caption{Normalised distributions of the top jet mass (top) and $\tau_{32}$ (bottom) for the single top signals and the dijet backgrounds, in the $1b$ sample with the baseline selection.}
\label{fig:dist1}
\end{center}
\end{figure}
\begin{figure}[t]
\begin{center}
\includegraphics[width=8cm]{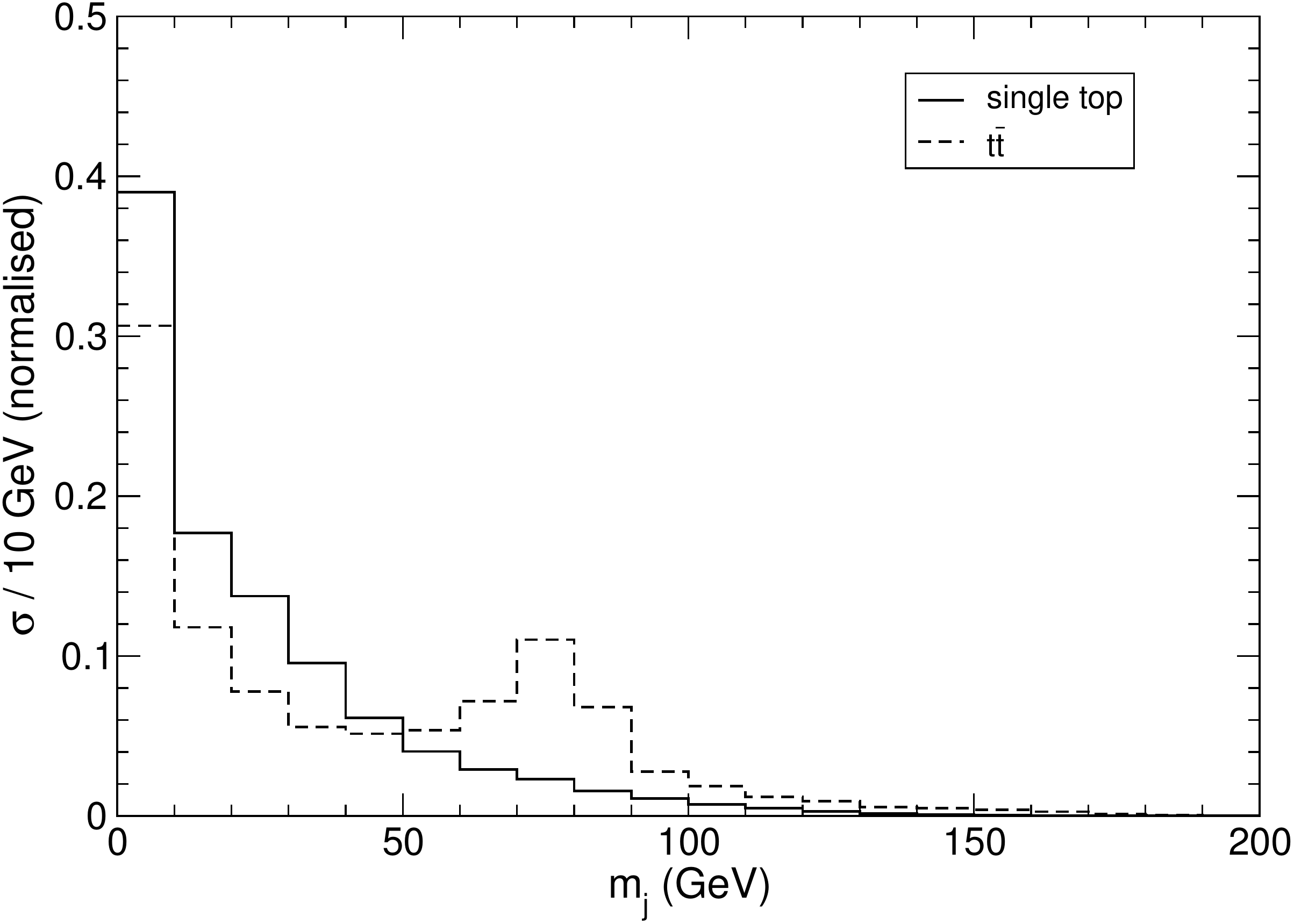}
\caption{Normalised distributions of the light jet mass for the single top signals and the $t \bar t$ background, in the $1b$ sample with the baseline selection.}
\label{fig:dist2}
\end{center}
\end{figure}

\begin{figure}[t]
\begin{center}
\begin{tabular}{c}
\includegraphics[width=7cm]{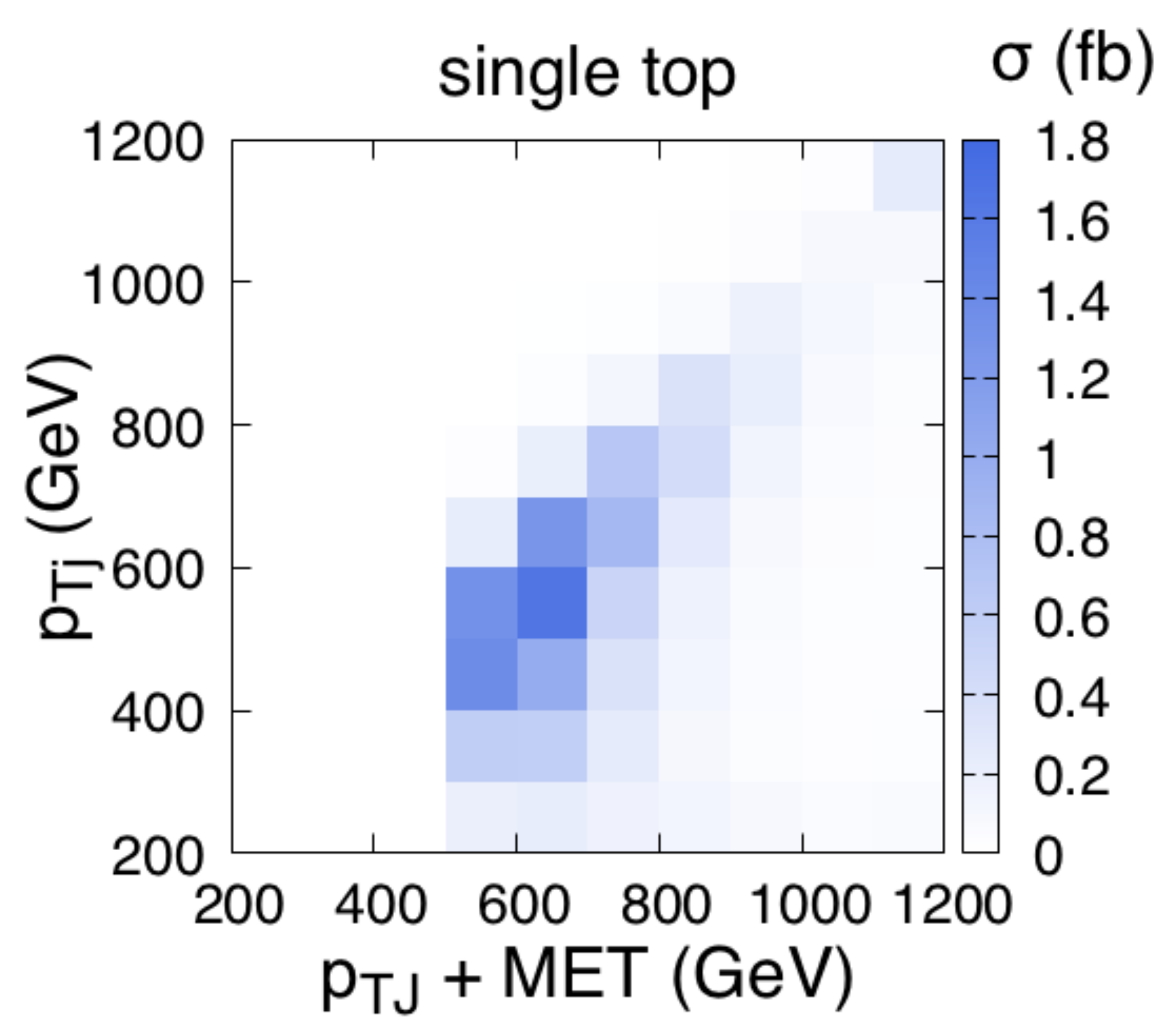} \\
\includegraphics[width=7cm]{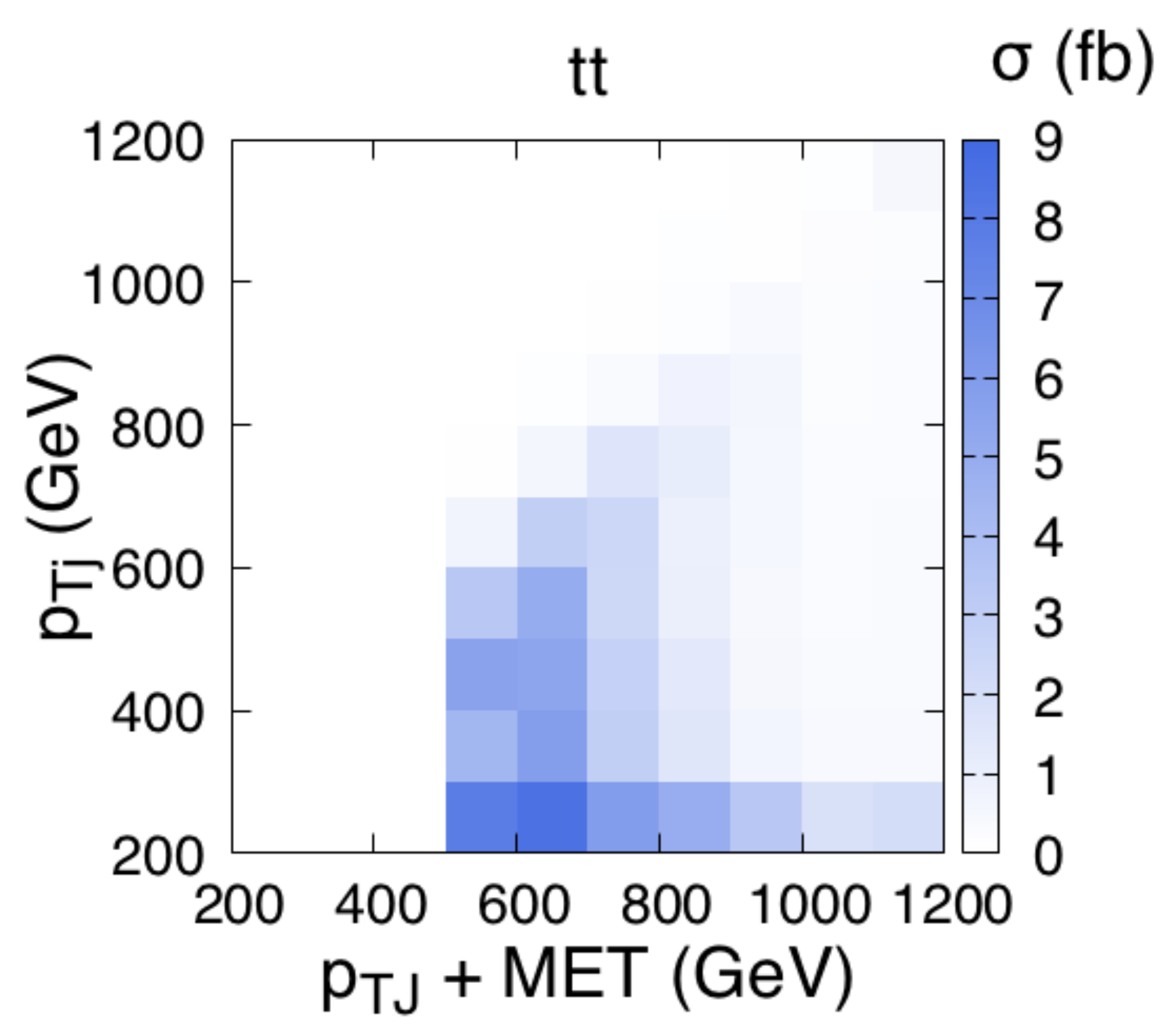}
\end{tabular}
\caption{Two-dimensional distributions of the sum of top jet $p_T$ plus missing transverse energy (MET) against the light jet $p_T$ for the single top signals (top) and the $t \bar t$ background, in the $1b$ sample with the baseline selection.}
\label{fig:dist3}
\end{center}
\end{figure}

\begin{figure}[htb]
\begin{center}
\begin{tabular}{c}
\includegraphics[width=8cm]{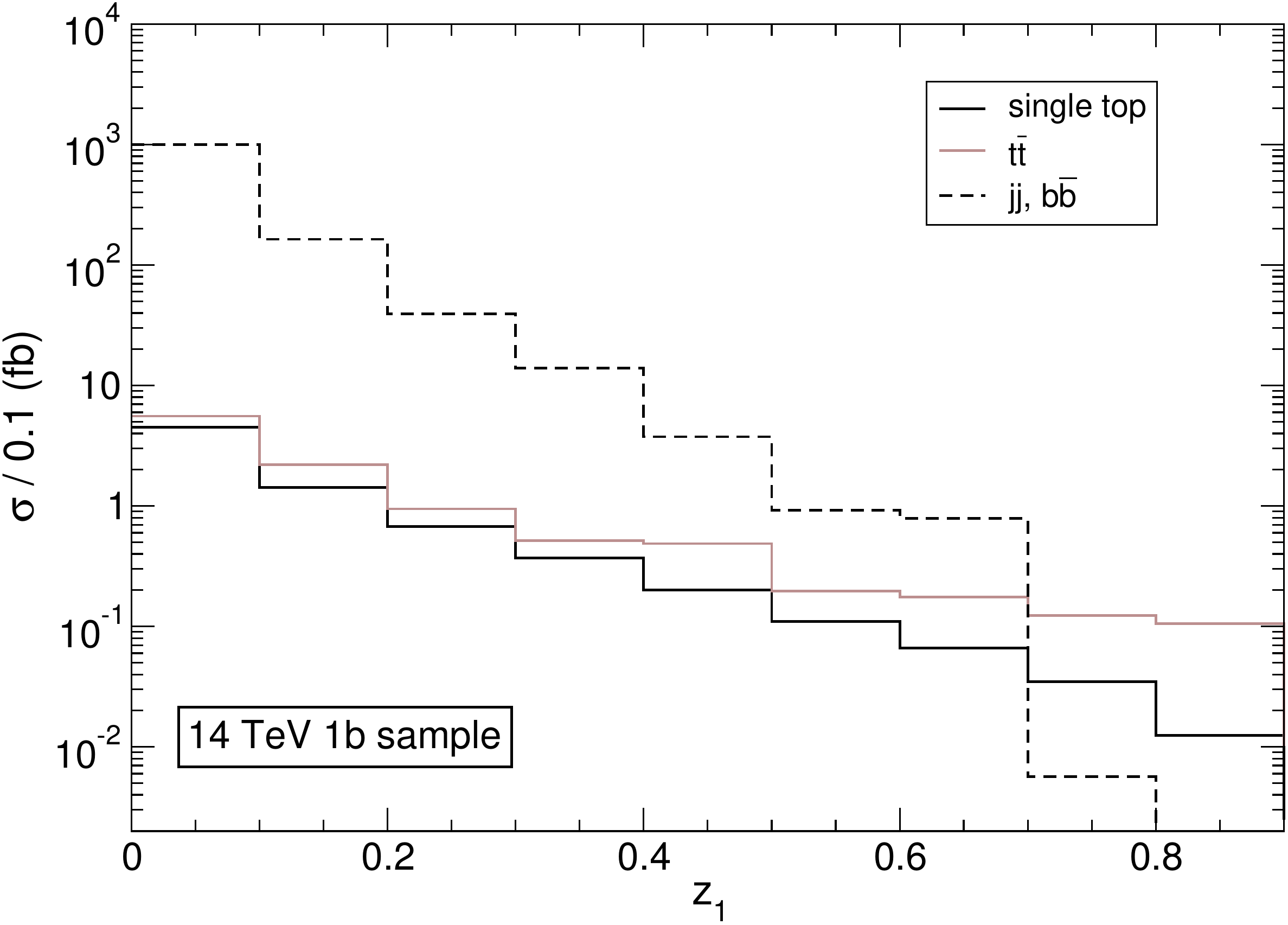} \\
\includegraphics[width=8cm]{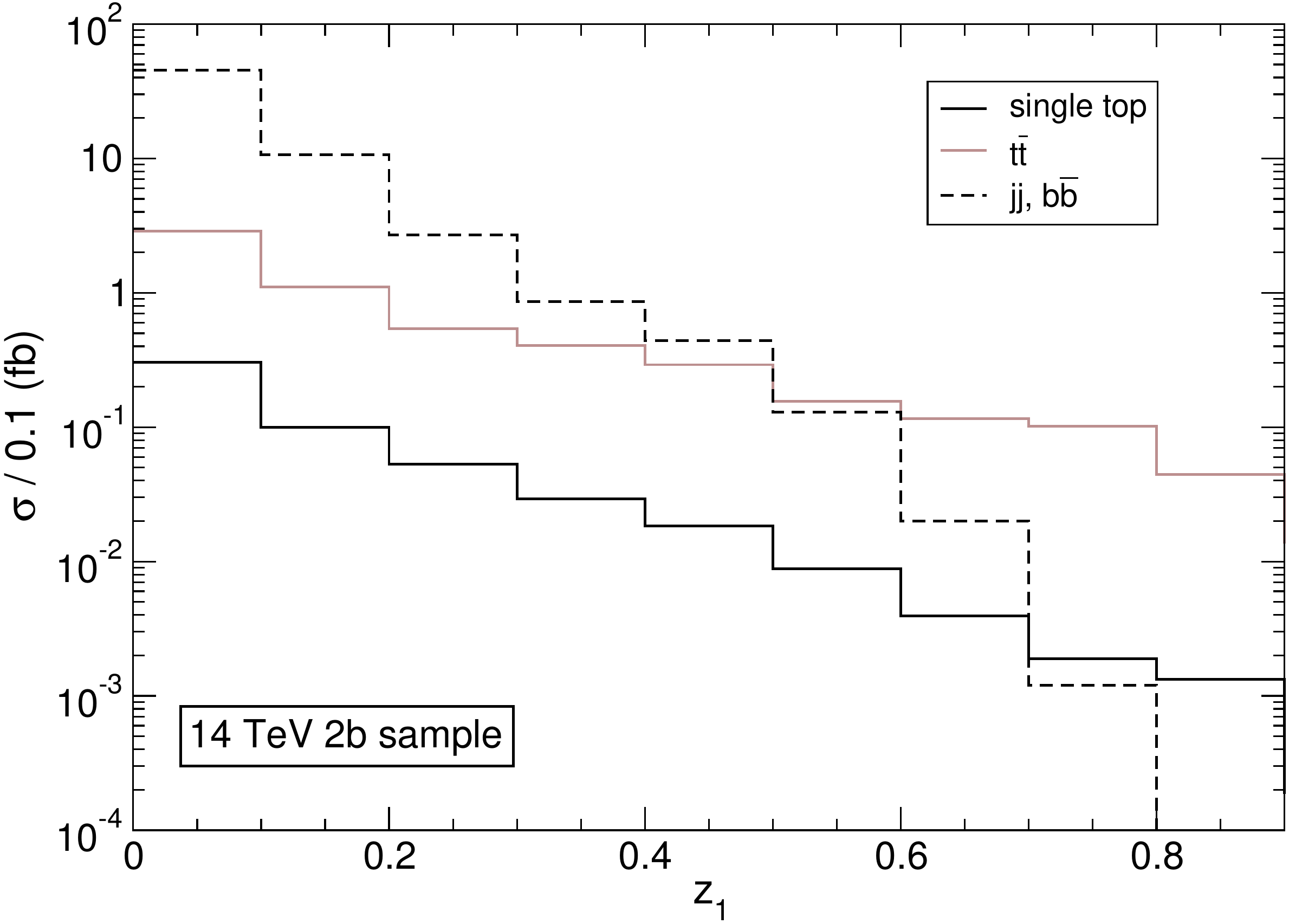}
\end{tabular}
\caption{Signal and background cross sections as a function of the lepton momentum fraction $z_1$ after the final selection in the $1b$ (top) and $2b$ (bottom) samples, for a CM energy of 14 TeV.}
\label{fig:dist4}
\end{center}
\end{figure}

Additional background suppression is achieved by exploiting kinematical differences between the signals and the backgrounds. The dijet ($jj$ and $b \bar b$) backgrounds can be reduced by considering the top jet mass $m_J$ and the sub-jettiness variable $\tau_{32} = \tau_3^{(1)} / \tau_2^{(1)}$~\cite{Thaler:2010tr}, shown in Fig.~\ref{fig:dist1}. We note that multivariate top taggers, even in simple setups~\cite{Komiske:2017aww}, have a better performance. However, for this exploratory work we will restrict ourselves to considering a simple substructure variable such as $\tau_{32}$.
The $t \bar t$ background can be suppressed by considering the light jet mass $m_j$, shown in Fig.~\ref{fig:dist2}, although $t \bar t$ events with small $m_j$ also result from the dilepton decay mode of the $t \bar t$ pair. In this case, one can use the balance of the jet momenta and missing transverse energy (MET). In the single-top signal, it is expected that for a boosted top quark decaying semileptonically its transverse momentum $p_{TJ}$ plus the missing energy will be approximately equal to the light jet transverse momentum $p_{Tj}$, as it is seen in Fig.~\ref{fig:dist3} (top). On the other hand, for dileptonic $t \bar t$ events the missing energy results from two neutrinos, so there is an imbalance, as seen in the bottom panel of Fig.~\ref{fig:dist3}.
We therefore use the following kinematical cuts:
\begin{itemize}
\item[(i)] $80 \leq m_J \leq 200$ GeV, and $\tau_{32} \leq 0.7$, aiming to reduce the dijet backgrounds.
\item[(ii)] $m_j \leq 60$ GeV and $p_{Tj} \geq 0.6 (p_{TJ} + \text{MET})$, in order to reduce the $t \bar t$ background.
\end{itemize}
The effect of these cuts on the signal and background cross sections is summarised in Table~\ref{tab:sig}.
Additional background reduction, especially of light dijet production, is achieved by requiring that the leading lepton $\ell_1$ has a large transverse momentum fraction $z_1 \equiv p_{Tl_1} / p_{TJ}$~\cite{Aguilar-Saavedra:2014iga}. The signal and background cross sections as a function of this variable are shown in Fig.~\ref{fig:dist4}. Note that in the $1b$ sample the distribution for the single top signals is steeper than for $t \bar t$. In the $tj$ process, dominant in this sample, the top quarks have a  polarisation $P_z \simeq 0.9$ in the direction of the jet $j$, which makes the leptons preferrably emitted opposite to the top quark direction (in the top quark rest frame). When boosted to the laboratory frame, the energy fraction $z_1$ is then typically smaller than in the unpolarised case. In the $2b$ sample the $tb$ process is dominant, with a polarisation $P_z \simeq 0.3$ in the direction of the $b$ quark. In this case, the effect of the polarisation in the $z_1$ distribution is milder, but still visible.

The further improvement of the signal significance based on this variable is discussed in section~\ref{sec:4}. We also considered applying a lepton veto near the light jet, to further suppress the $t \bar t$ background in the dilepton decay channel, but found no significant improvement.

\section{Analysis at HE-LHC}
\label{sec:3}

For the study at 27 TeV we follow the same steps described in the previous section, generating samples in the same $p_T$ intervals and with the same Monte Carlo statistics. Because the kinematics in the high-$p_T$ range is similar at 14 and 27~TeV, the main difference being the cross section increase at 27 TeV, we keep the same event selection for simplicity. The cross sections for the different processes with the baseline selection, after the separate sets of cuts (i) and (ii), and after the final selection, are collected in table~\ref{tab:sig2}. The luminosity assumed is $\mathcal{L} = 15~\text{ab}^{-1}$.

\begin{table}[htb]
\begin{center}
\begin{tabular}{ccccccccc}
& \multicolumn{2}{c}{baseline} & \multicolumn{2}{c}{(i) only}  & \multicolumn{2}{c}{(ii) only} & \multicolumn{2}{c}{(i) $+$ (ii)} \\
& $1b$ & $2b$ & $1b$ & $2b$ & $1b$ & $2b$ & $1b$ & $2b$ \\
$tj$ [fb] & 81.5 & 3.22
       & 51.9 & 1.98
       & 59.8 & 1.40
       & 38.0 & 0.875
 \\
$tb$ [fb] & 2.29 & 3.55
        & 1.19 & 2.00 
        & 1.19 & 1.40 
        & 0.679 & 1.31 
\\
$t \bar t$ [fb] & 685 & 425
               & 468 & 286
               & 107 & 54.5
               & 71.0 & 35.6
\\
$b \bar b$ [fb] & 554 & 964
                 & 86.3 & 155
                 & 340 & 653
                 & 53.1 & 103
\\
$jj$ [pb] & 44.9 & 1.74
       & 11.3 & 0.518
       & 27.6 & 0.939
       & 6.79 & 0.312
\\
\end{tabular}
\caption{Cross sections for signals and backgrounds at different stages of event selection, for a CM energy of 27 TeV.}
\label{tab:sig2}
\end{center}
\end{table}

\begin{figure}[htb]
\begin{center}
\begin{tabular}{c}
\includegraphics[width=8cm]{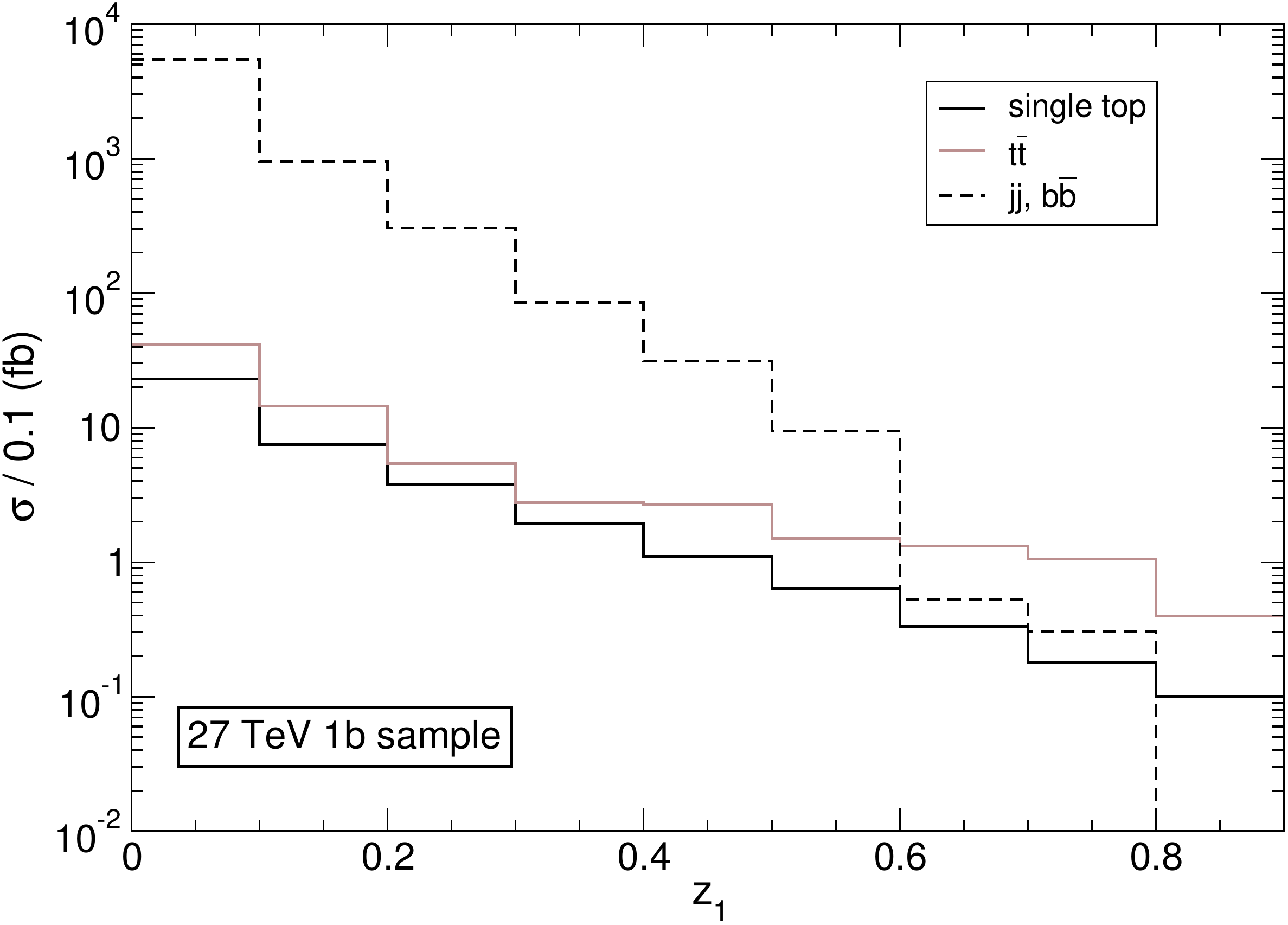} \\
\includegraphics[width=8cm]{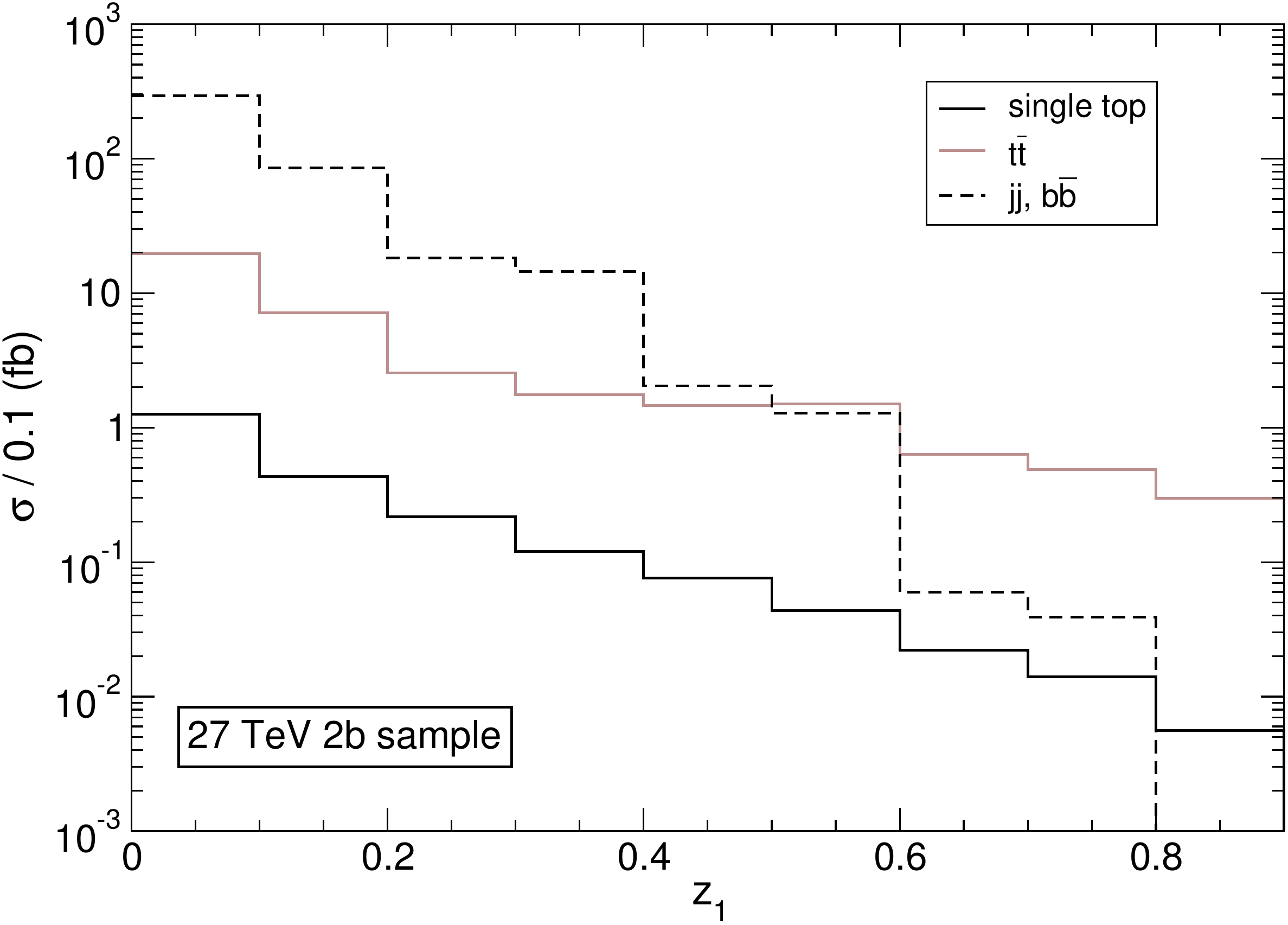}
\end{tabular}
\caption{Signal and background cross sections as a function of the lepton momentum fraction $z_1$ after the final selection in the $1b$ (top) and $2b$ (bottom) samples, for a CM energy of 27 TeV.}
\label{fig:dist5}
\end{center}
\end{figure}

We observe that the $tj$, $t \bar t$, $b\bar b$ and $jj$ cross sections increase by a factor of ten, while the $tb$ cross section increases by a smaller factor around six. The signal and background cross sections as a function of the lepton momentum fraction $z_1$ are presented in Fig.~\ref{fig:dist5}.

\section{Analysis at FCC}
\label{sec:3b}

\begin{figure}[t]
\begin{center}
\begin{tabular}{c}
\includegraphics[width=8cm]{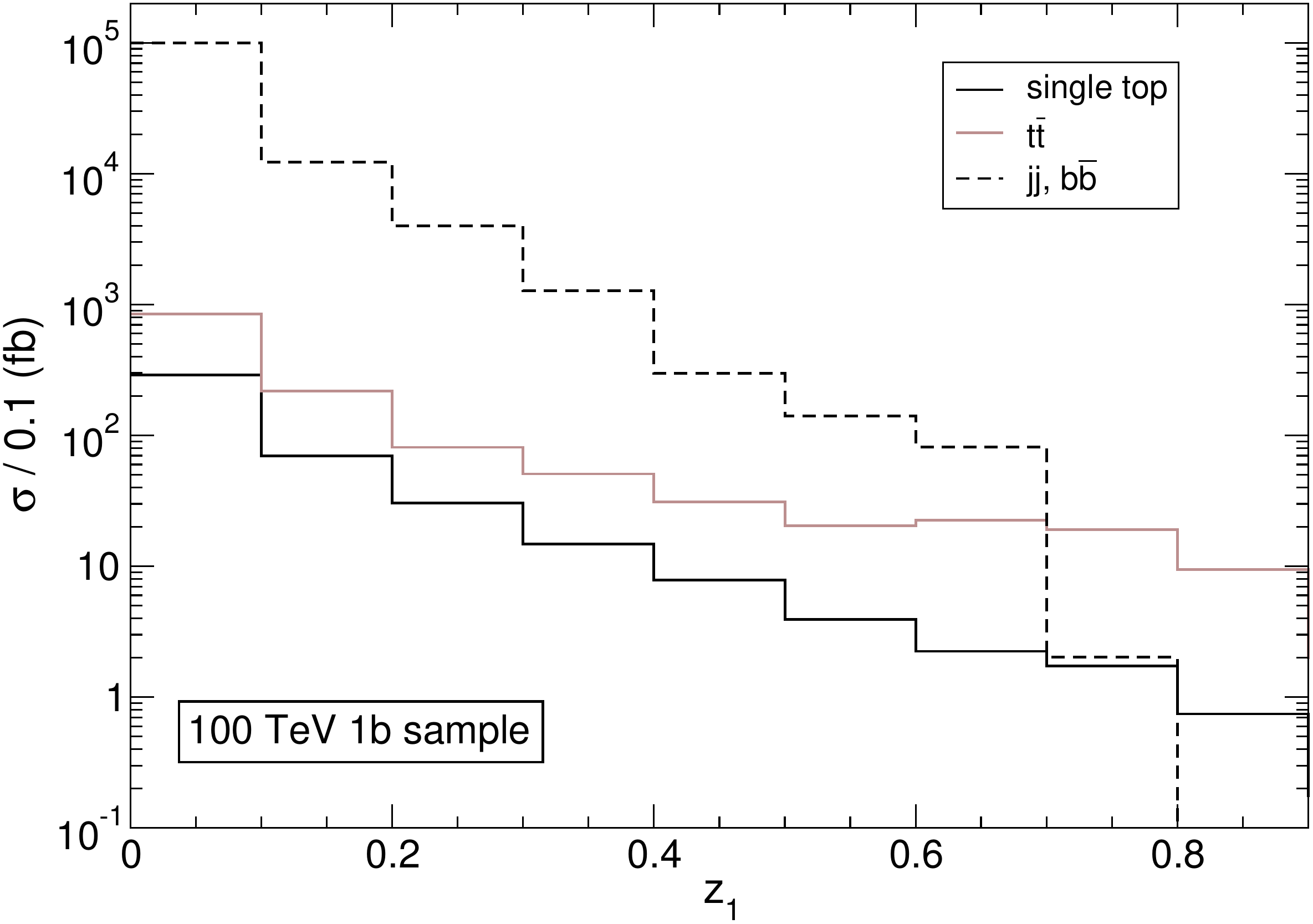} \\
\includegraphics[width=8cm]{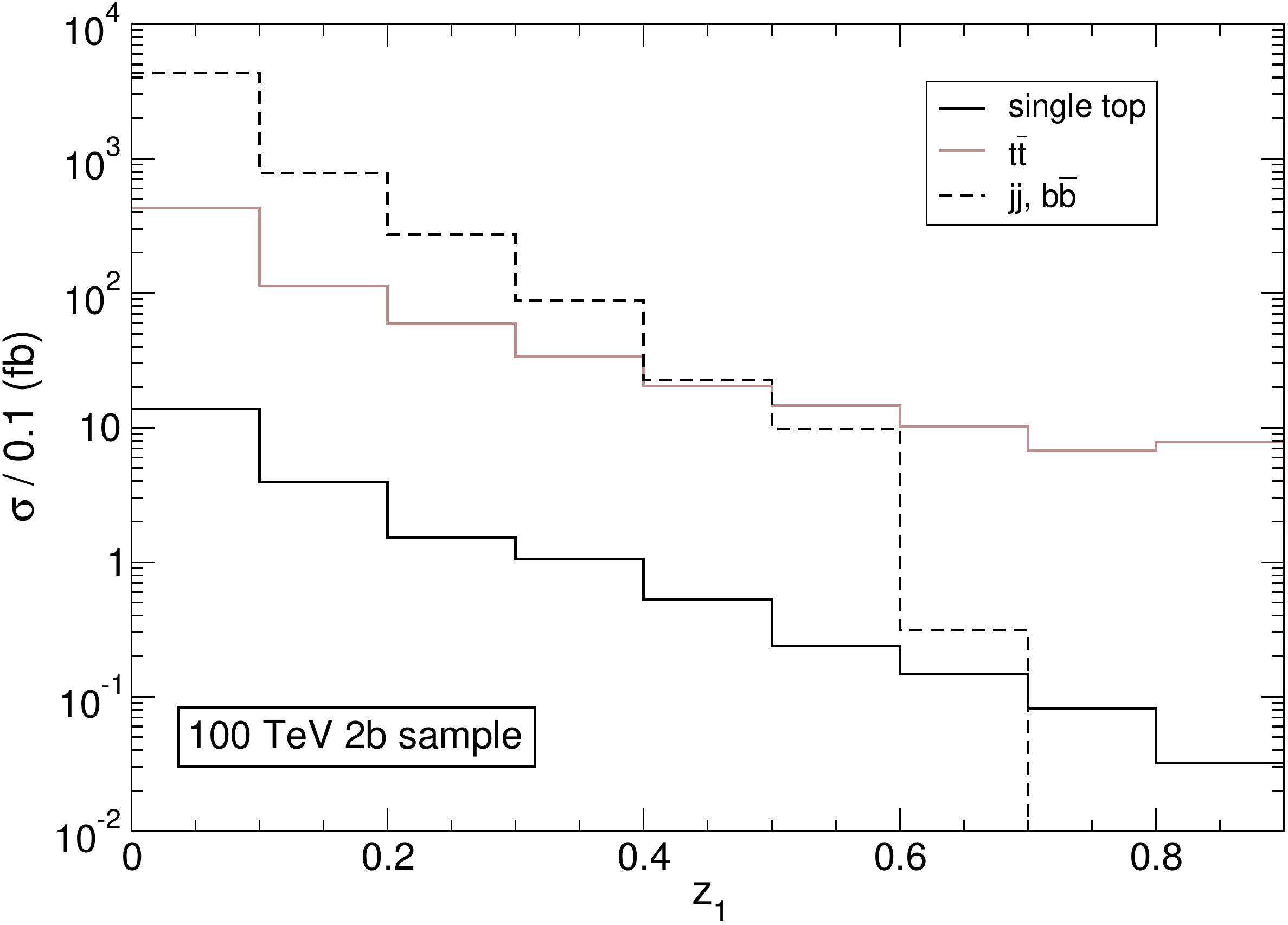}
\end{tabular}
\caption{Signal and background cross sections as a function of the lepton momentum fraction $z_1$ after the final selection in the $1b$ (top) and $2b$ (bottom) samples, for a CM energy of 100 TeV.}
\label{fig:dist6}
\end{center}
\end{figure}

The Monte Carlo event generation and simulation at 100~TeV proceeds in the same way as described in Section~\ref{sec:2}, but using the Delphes card for the FCC~\cite{Benedikt:2018csr}. 
We have explored raising the lower cut $p_{TJ} \geq 500$ GeV but we find no improvement in the sensitivity, therefore the event selection is kept the same as for lower CM energies. 
The cross sections for the different processes with the baseline selection, after the separate sets of cuts (i) and (ii), and after the final selection, are collected in table~\ref{tab:sig3}. The luminosity assumed is $\mathcal{L} = 30~\text{ab}^{-1}$. The signal and background cross sections as a function of the lepton momentum fraction $z_1$ are presented in Fig.~\ref{fig:dist6}. We note that the $tj$ and $tb$ single top cross sections increase by factors of 70 and 30 with respect to the HL-LHC energy, while the increase in the background is larger, by factors of $130-160$. However, the overall increase in statistics allows to perform measurements with a higher precision, as seen in the following.

\begin{table}[t]
\begin{center}
\begin{tabular}{ccccccccc}
& \multicolumn{2}{c}{baseline} & \multicolumn{2}{c}{(i) only}  & \multicolumn{2}{c}{(ii) only} & \multicolumn{2}{c}{(i) $+$ (ii)} \\
& $1b$ & $2b$ & $1b$ & $2b$ & $1b$ & $2b$ & $1b$ & $2b$ \\
$tj$ [fb] & 1020 & 51.2
             & 583 & 25.7
             & 733 & 22.1
             & 417 & 12.4
 \\
$tb$ [fb] & 17.8 & 27.2
              & 8.31 & 13.5 
              & 9.38 & 16.4 
              & 5.04 & 8.88 
\\
$t \bar t$ [pb] & $15.5$ & 9.36
                      & 9.58 & 5.75
                      & 2.25 & 1.19
                      & 1.30 & 0.697
\\
$b \bar b$ [pb] & $12.4$ & $20.8$
                        & 1.42 & 2.480
                        & 7.39 & $14.1$
                        & 0.854 & 1.61
\\
$jj$ [pb] & 999 & 42.0
             & 203 & 9.15
             & 578 & 20.1
             & 117 & 3.90
\\
\end{tabular}
\caption{Cross sections for signals and backgrounds at different stages of event selection, for a CM energy of 100 TeV.}
\label{tab:sig3}
\end{center}
\end{table}

\section{Limits on anomalous couplings}
\label{sec:4}

\begin{figure}[t]
\begin{center}
\begin{tabular}{c}
\includegraphics[width=8cm]{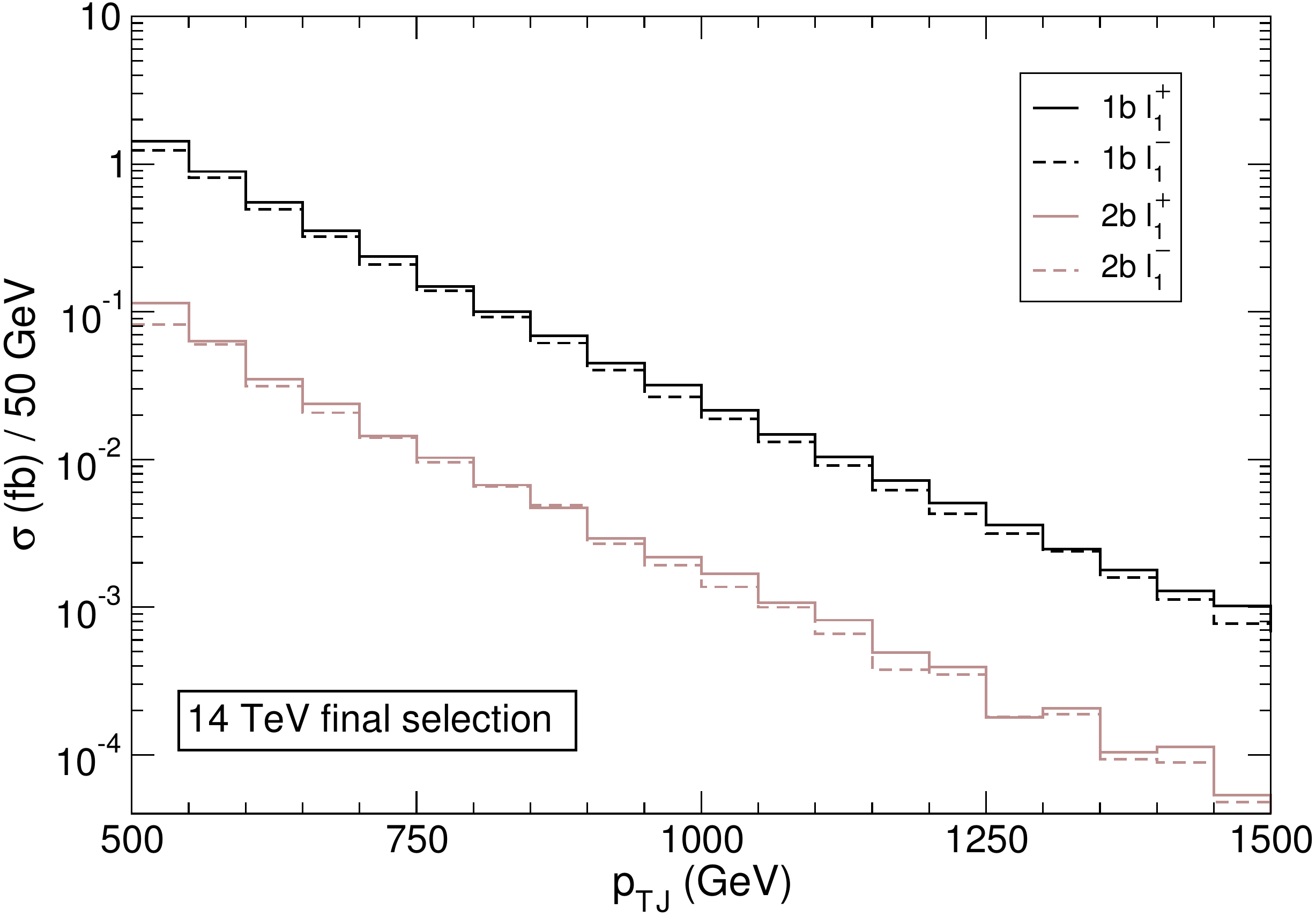} \\
\includegraphics[width=8cm]{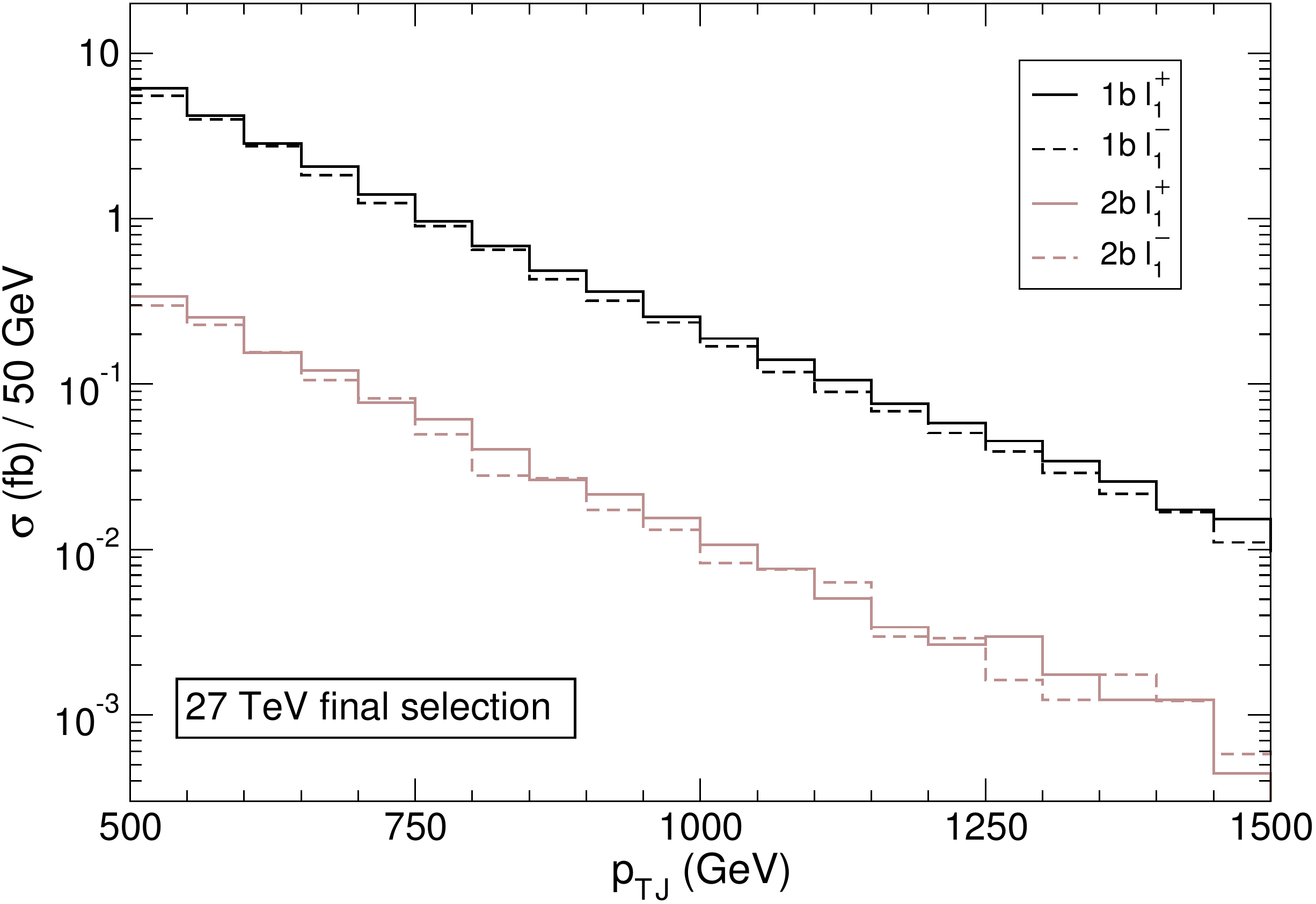} \\
\includegraphics[width=8cm]{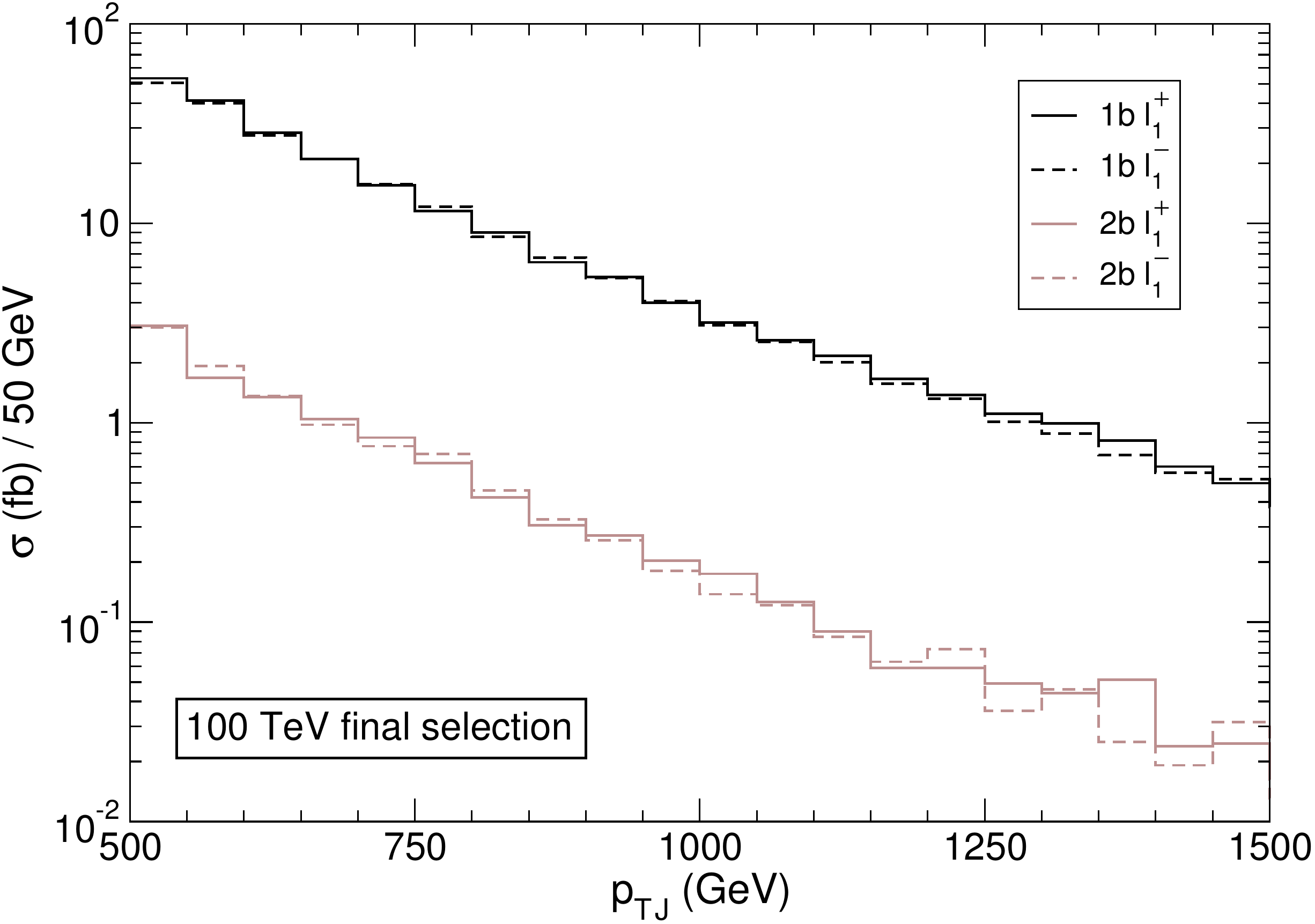}
\end{tabular}
\caption{Signal cross sections for positive and negative charged leptons }
\label{fig:dist7}
\end{center}
\end{figure}

\begin{figure}[t]
\begin{center}
\begin{tabular}{c}
\includegraphics[width=8cm]{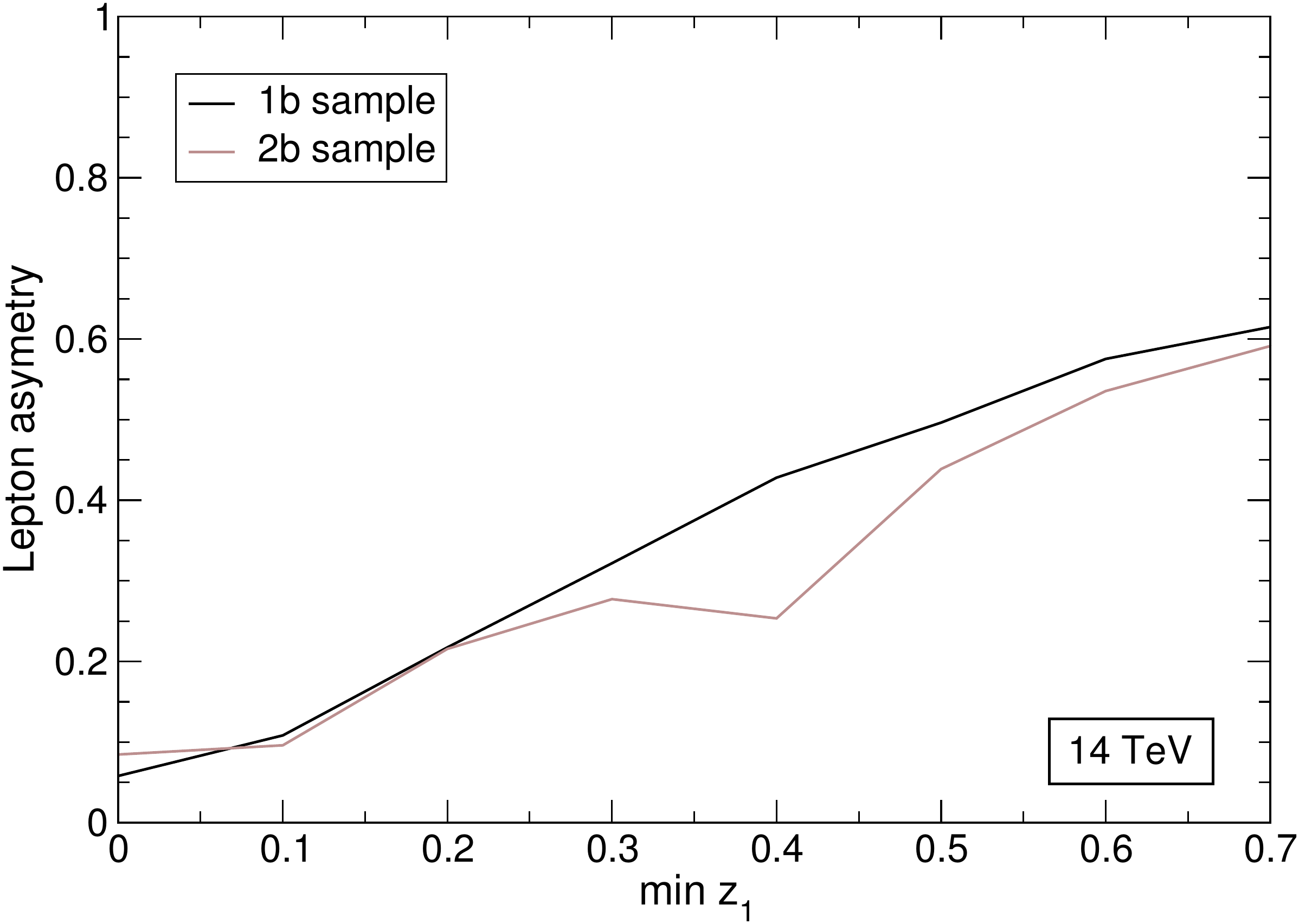} \\
\includegraphics[width=8cm]{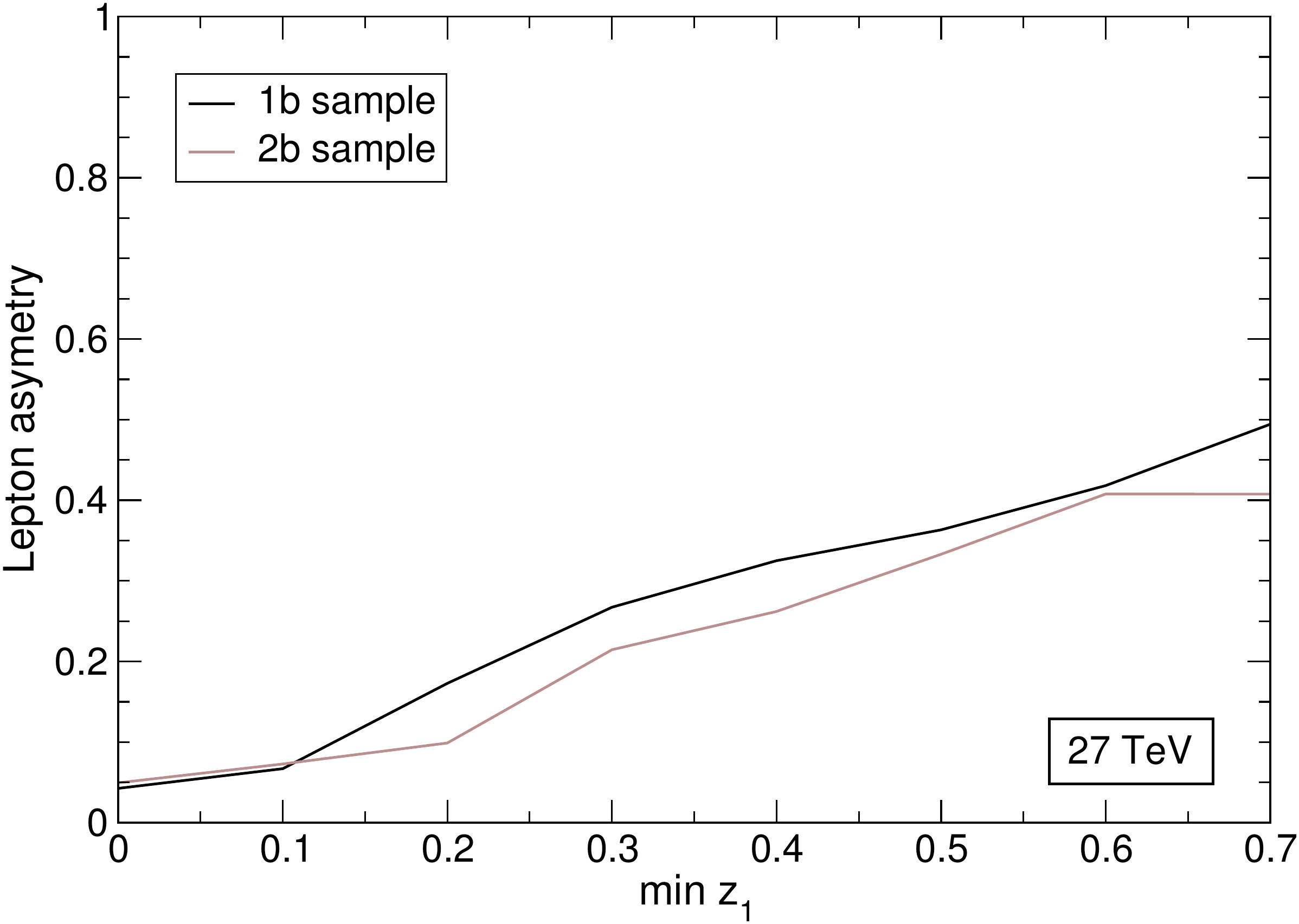} \\
\includegraphics[width=8cm]{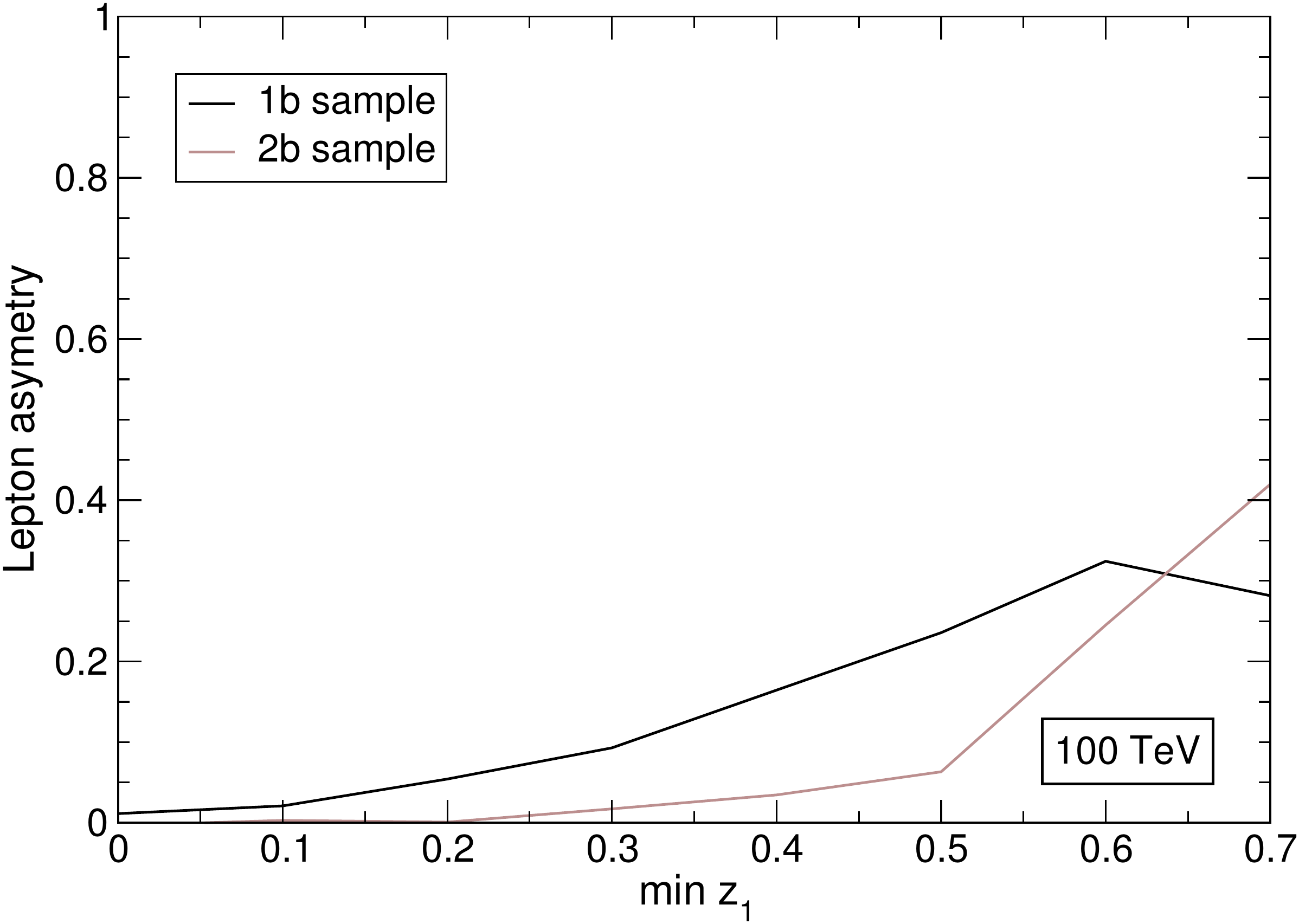}
\end{tabular}
\caption{Lepton charge asymmetry in Eq.~(\ref{ec:asym}) as a function of the lower cut on $z_1$.}
\label{fig:asym}
\end{center}
\end{figure}

The event selection discussed in the previous sections allows to significantly increase the signal to background ratio, for example from $6.9 \times 10^{-4}$ with topology cuts at 14 TeV, to $1.9\;(3.0) \times 10^{-3}$ with the baseline selection in the $1b$ ($2b$) samples,  and $6.0\;(7.9) \times 10^{-3}$ with the final event selection at the same energy. Still, the single top signals are too small to be seen without a precise normalisation of the background. In order to do this, one can exploit the fact that the signals have larger cross sections for top quark production than for antiquarks, leading to more events where the leading lepton $\ell_1$ is positive, compared to events where $\ell_1$ is negative. This can be seen in Fig.~\ref{fig:dist7}, where signal cross sections for positive and negative $\ell_1$ are presented.
On the other hand, the backgrounds are charge symmetric. 
We can define a lepton asymmetry
\begin{equation}
A_\ell = \frac{\sigma^+ - \sigma^-}{\sigma^+ + \sigma^-} \,,
\label{ec:asym}
\end{equation}
where $\sigma^\pm$ refers to the signal cross sections where the leading lepton, required to be within $\Delta R \leq 0.5$ of the top jet as already mentioned, is positive or negative. 
The asymmetry as a function of the lower cut on $z_1$ is shown in Fig.~\ref{fig:asym}. 
We point out that $A_\ell$ is washed out with respect to the corresponding asymmetry of $t$ versus $\bar t$ production because charged leptons of either sign also result from $b$ quark decays. 
Therefore, raising the lower cut on $z_1$ increases the asymmetry, as the contributions from $b$ quark decays are suppressed. (The asymmetry is smaller at 100 TeV because the symmetric background grows more quickly with energy, as mentioned before.) Additionally, requiring higher $z_1$ suppresses dijet and $b \bar b$ backgrounds, making $t \bar t$ the dominant one. We collect in Table~\ref{tab:sig4} the cross sections for $z_1 \geq 0.6$, in the $1b$ and $2b$ samples.

\begin{table}[htb]
\begin{center}
\begin{tabular}{lcccccc}
& \multicolumn{2}{c}{14 TeV} & \multicolumn{2}{c}{27 TeV} & \multicolumn{2}{c}{100 TeV}  \\
& $1b$ & $2b$ & $1b$ & $2b$ & $1b$ & $2b$ \\
$tj$ & 0.113 & $1.98 \times 10^{-3}$ & 0.610 & 0.0135 & 4.80 & 0.142 \\
$tb$ & $3.37 \times 10^{-3}$ & $6.24 \times 10^{-3}$ & 0.0134 & 0.0242 & 0.0778 & 0.134 \\
$t \bar t$ & 0.432 & 0.275 & 2.78 & 1.81 & 53.9 & 26.3 \\
$b \bar b$ & 0.011 & 0.020 & 0.0725 & 0.131 & 0.333 & 0.570 \\
$jj$ & 0.23 & $6.9 \times 10^{-3}$ & 0.92 & 0.037 & 18 & 0.68 \\
\end{tabular}
\caption{Cross sections (in fb) for the different processes in the $1b$ and $2b$ samples with the final selection, for $z_1 \geq 0.6$. }
\label{tab:sig4}
\end{center}
\end{table}

Taking advantage of this asymmetry in the signal, one can use the number of (signal) events with positive leptons minus the number of events with negative leptons, 
\begin{equation}
\Delta = S^+ - S^- \,, \quad S^\pm = \mathcal{L} \times \sigma^\pm \,,
\end{equation}
in order to set limits on possible new physics contributions.\footnote{We have also explored the ratio $(S^+ - S^-)/(S^+ + S^-)$, but the sensitivity of the ratio is smaller due to the large (and uncorrelated to the numerator) scale uncertainty in the denominator, dominated by the $jj$ and $t \bar t$ processes.}
Including a (relative) systematic uncertainty $\eta$ on the SM prediction for the difference $\Delta$, its expected significance is
\begin{equation}
n_\sigma = \frac{\Delta}{\sqrt{B + (\eta \Delta)^2}} \,.
\end{equation}
We assume $\eta = 0.1$ for our estimations of the sensitivity for HL-LHC and HE-LHC. This assumption is based on the extrapolation of the values for current total cross section measurements~\cite{Sirunyan:2018rlu} assuming that the uncertainty from Monte Carlo modeling is halved. Note however that
\begin{enumerate}
\item[(i)] a direct measurement of the cross section in Drell-Yan processes $pp \to W^\pm \to \ell^\pm \nu$ can be used to predict the $s$-channel single (anti-)top cross sections, which dominate the $2b$ samples;
\item[(ii)] the uncertainty in the $t$-channel single top cross section is fully correlated in the $1b$ and $2b$ samples, and the combination of both measurements would have smaller systematics. Alternatively, one can use the measurement in one of the final states to predict the $tj$ cross section in the other one.
\end{enumerate}
At 14 TeV, in the $1b$ and $2b$ subsamples the significance is nearly maximal for
\begin{align}
& 1b: \quad z_1 \geq 0.5 \quad\rightarrow\quad n_\sigma = 3.7 \;[4.0] \,, \notag \\
& 2b: \quad z_1 \geq 0.3 \quad\rightarrow\quad n_\sigma = 0.6 \;[0.6] \,.
\end{align}
At 27 TeV we have
\begin{align}
& 1b: \quad z_1 \geq 0.6 \quad\rightarrow\quad n_\sigma = 8.6 \;[16.8]\,, \notag \\
& 2b: \quad z_1 \geq 0.4 \quad\rightarrow\quad n_\sigma = 1.8 \;[1.8] \,.
\end{align}
The numbers in brackets correspond to the significances in the absence of systematic uncertainties, that is, for $\eta = 0$. For FCC, we take two assumptions on systematic uncertainties: a conservative estimate $\eta = 0.1$, as taken for the LHC upgrades, and a more optimistic one $\eta = 0.01$. In the former case we find
\begin{align}
& 1b: \quad z_1 \geq 0.6 \quad\rightarrow\quad n_\sigma = 9.2 \;[23.5] \,, \notag \\
& 2b: \quad z_1 \geq 0.6 \quad\rightarrow\quad n_\sigma = 2.2 \;[2.2] \,.
\end{align}
Again, the numbers between brackets are the significances for $\eta = 0$. For $\eta = 0.01$ the impact of systematics is very small,
\begin{align}
& 1b: \quad z_1 \geq 0.6 \quad\rightarrow\quad n_\sigma = 22.8  \,, \notag \\
& 2b: \quad z_1 \geq 0.6 \quad\rightarrow\quad n_\sigma = 2.2  \,.
\end{align}

Although some of these significances are modest, they lead to competitive constraints on possible $tbW$ anomalous couplings, as the cross section enhancement in the presence of such anomalous contributions would be huge at high momenta. For illustration, we have considered an anomalous interaction of the type  
\begin{equation}
-\frac{g}{\sqrt 2 M_W} g_L \bar b_R \sigma^{\mu \nu} t_L \partial_\mu W_\nu^- + \text{H.c.} \,,
\label{ec:lagr}
\end{equation}
in standard notation, with $g$ the electroweak coupling and $M_W$ the $W$ boson mass. We have calculated the single $t$ and single $\bar t$ cross sections at 14, 27 and 100 TeV in the presence of such term, in each of the $p_T$ bins used for our simulation, by using {\scshape Protos}~\cite{AguilarSaavedra:2008gt}. (This is done by computing the cross sections, bin by bin, for five different values of the anomalous coupling $g_L$, to subsequently obtain the analytical dependence with a fit.) For example, for top transverse momentum $p_T \in [800,900]$ GeV at the parton level, the $t$-channel cross sections at 14 TeV are
\begin{align}
& \sigma(t) = 5.3 + 620 \, |g_L|^2~\text{fb} \,, \notag \\
& \sigma(\bar t) = 1.3 + 270 \, |g_L|^2~\text{fb} \,,
\end{align}
omitting small interference terms. The cross section enhancement in the presence of anomalous interactions is manifest. At 27 and 100 TeV the cross sections are larger, but the relative enhancement with respect to the SM value due to the presence of anomalous interactions remains almost the same, within each $p_T$ bin, as it depends on the momenta in the partonic subprocess. We obtain at 14 TeV the upper limits at 95\% confidence level (CL)
\begin{align}
& 1b: \quad  |g_L| \leq 0.087 \; [0.062] \,, \notag \\
& 2b: \quad |g_L| \leq 0.114 \; [0.082] \,.
\label{ec:lim14}
\end{align}
For completeness, limits at one standard deviation ($1 \sigma$) are given between brackets.
At 27 TeV the expected limits are
\begin{align}
& 1b: \quad  |g_L| \leq 0.050 \; [0.036] \,, \notag \\
& 2b: \quad |g_L| \leq 0.068  \; [0.049] \,.
\label{ec:lim27}
\end{align}
At 100 TeV, assuming 10\% systematic uncertainties the expected limits are
\begin{align}
& 1b: \quad  |g_L| \leq 0.046 \; [0.033] \,, \notag \\
& 2b: \quad |g_L| \leq 0.043  \; [0.031] \,.
\label{ec:lim100a}
\end{align}
Note that the sensitivity of the best final state ($1b$) is dominated by systematics already at 27 TeV, therefore the improvement brought by the FCC energy increase and larger statistics is marginal. On the other hand, in the $2b$ final state the impact of systematic uncertainties is still small. If the systematic uncertainties can be reduced to 1\%, the corresponding limits at 100 TeV are
\begin{align}
& 1b: \quad  |g_L| \leq 0.030 \; [0.022] \,, \notag \\
& 2b: \quad |g_L| \leq 0.043  \; [0.031] \,.
\label{ec:lim100b}
\end{align}

We note that in the presence of a non-zero $g_L$ the charged lepton distribution in the top quark rest frame does not change, even at quadratic order~\cite{AguilarSaavedra:2010nx}, therefore it is justified to ignore the effect of the anomalous interaction in the top quark decay, in particular in the distribution of the lepton momentum fraction $z_1$. We have verified that the effects in the top quark polarisation for $g_L =O(0.1)$ are at the permille level, therefore the effect of the non-zero anomalous coupling in the signal acceptance is well described by our computation of the cross section scaling factors in narrow slices of $p_T$.

\section{Limits on weakly-coupled $W'$ bosons}
\label{sec:5}

In the search for $W' \to tb$ we consider a leptophobic $W'$ boson with right-handed couplings given by the Lagrangian
\begin{eqnarray}
\mathcal{L}_{W'} & = & - \frac{g_{W'}}{\sqrt 2} \left( 
\bar d_R \gamma^\mu u_R + \bar s_R \gamma^\mu c_R  + \bar b_R \gamma^\mu t_R  
\right) W_\mu^- \notag \\
& & + \text{H.c.}
\end{eqnarray}
We choose right-handed couplings because a $W'$ coupling to left-handed fermions will generally couple to leptons~\cite{delAguila:2010mx}, thereby producing clean leptonic signals. Limits on leptophobic $W'$ bosons arise from  their decay into $tb$ final states~\cite{Sirunyan:2017vkm,Aaboud:2018jux}. 
For a coupling $g_{W'} = g$, masses up to $M_{W'} = 3.6$ TeV are excluded at the 95\% CL.

The production and decay process $pp \to W' \to tb$ does not interfere with $s$-channel single top in the limit of massless $u$, $d$, $c$, $s$ quarks. As benchmark examples we consider three masses, $2$, $3$ and $4$ TeV. As mentioned before, we first focus on the sensitivity to very weakly-coupled intermediate mass $W'$ bosons. We shall return later to the study of the ultimate mass reach at 100 TeV for $W'$ bosons with $g_{W'} \simeq g$.
In the event generation we set $g_{W'} = 0.1$, for which the total widths
are $\Gamma_{W'} = 1.2$, $1.8$ and $2.4$ GeV for $M_{W'} = 2$, $3$ and $4$ TeV, respectively. These widths are much smaller than the experimental mass resolution, therefore the results for different values of the coupling can simply be obtained by scaling the total cross section.
Samples of $6 \times 10^4$ events are generated for each $W'$ mass and collider energy, including all the decay channels of the top quark.

We restrict our event selection to the $2b$ final state, because a $b$ quark is already present in the $W' \to tb$ decay. 
In contrast with the non-resonant anomalous coupling signals studied in the previous section, a new $W'$ boson can be detected via the presence of a bump in the reconstructed $W'$ invariant mass distribution.
As a proxy for the $W'$ mass we use the invariant mass of the two jets plus the neutrino, $m_{Jj\nu}$. The transverse component of the neutrino momentum is set to the missing energy in the event, and the longitudinal component (and energy) are determined by requiring that the invariant mass of the neutrino and the leading charged lepton equal the $W$ boson mass. This constraint yields a second degree equation; among the two solutions we choose the one that gives smaller longitudinal momentum. When the equation does not have solution we determine the longitudinal momentum by setting the neutrino three-momentum parallel to the top jet three-momentum.

\begin{figure}[t]
\begin{center}
\begin{tabular}{c}
\includegraphics[width=8cm]{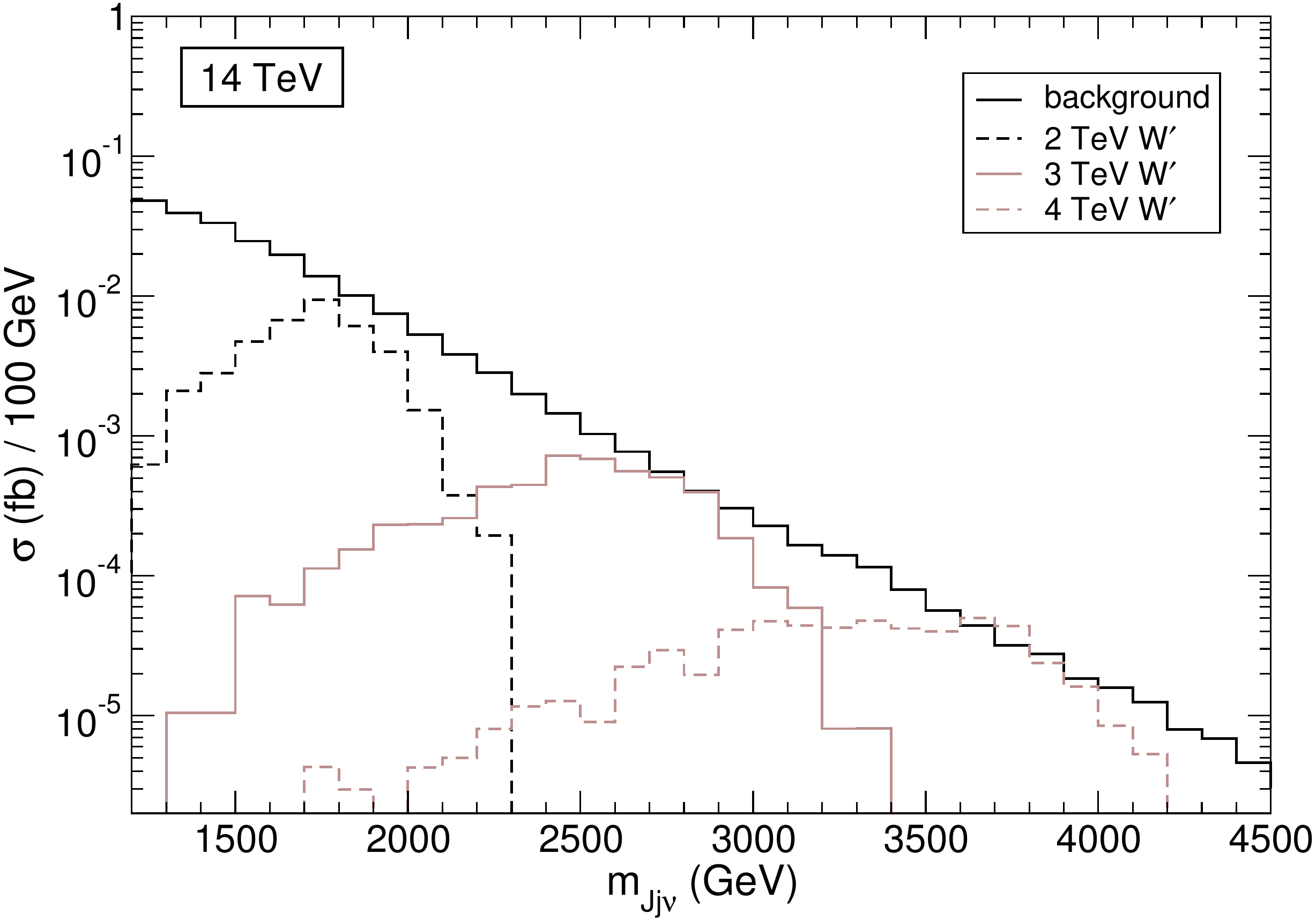} \\
\includegraphics[width=8cm]{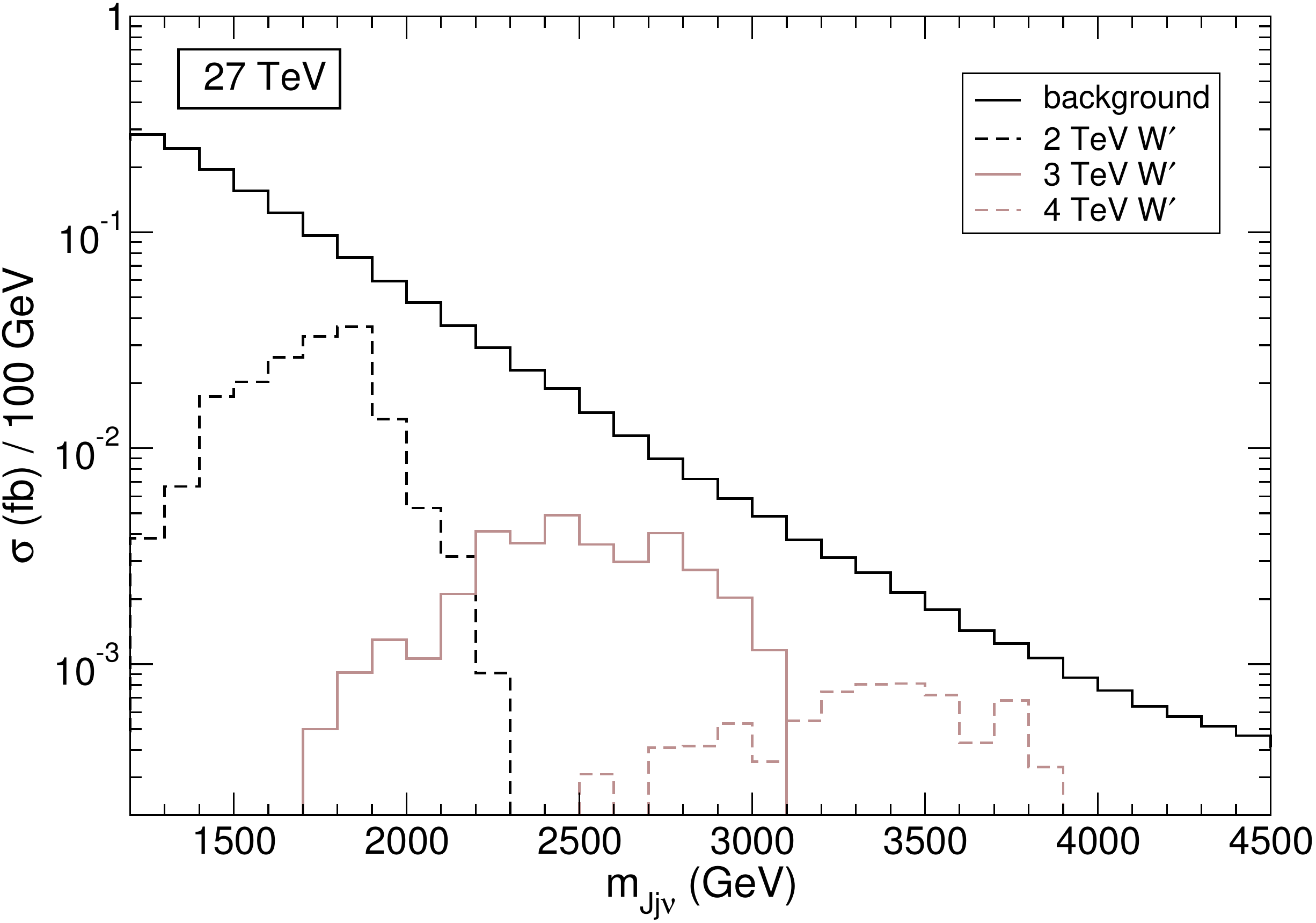} \\
\includegraphics[width=8cm]{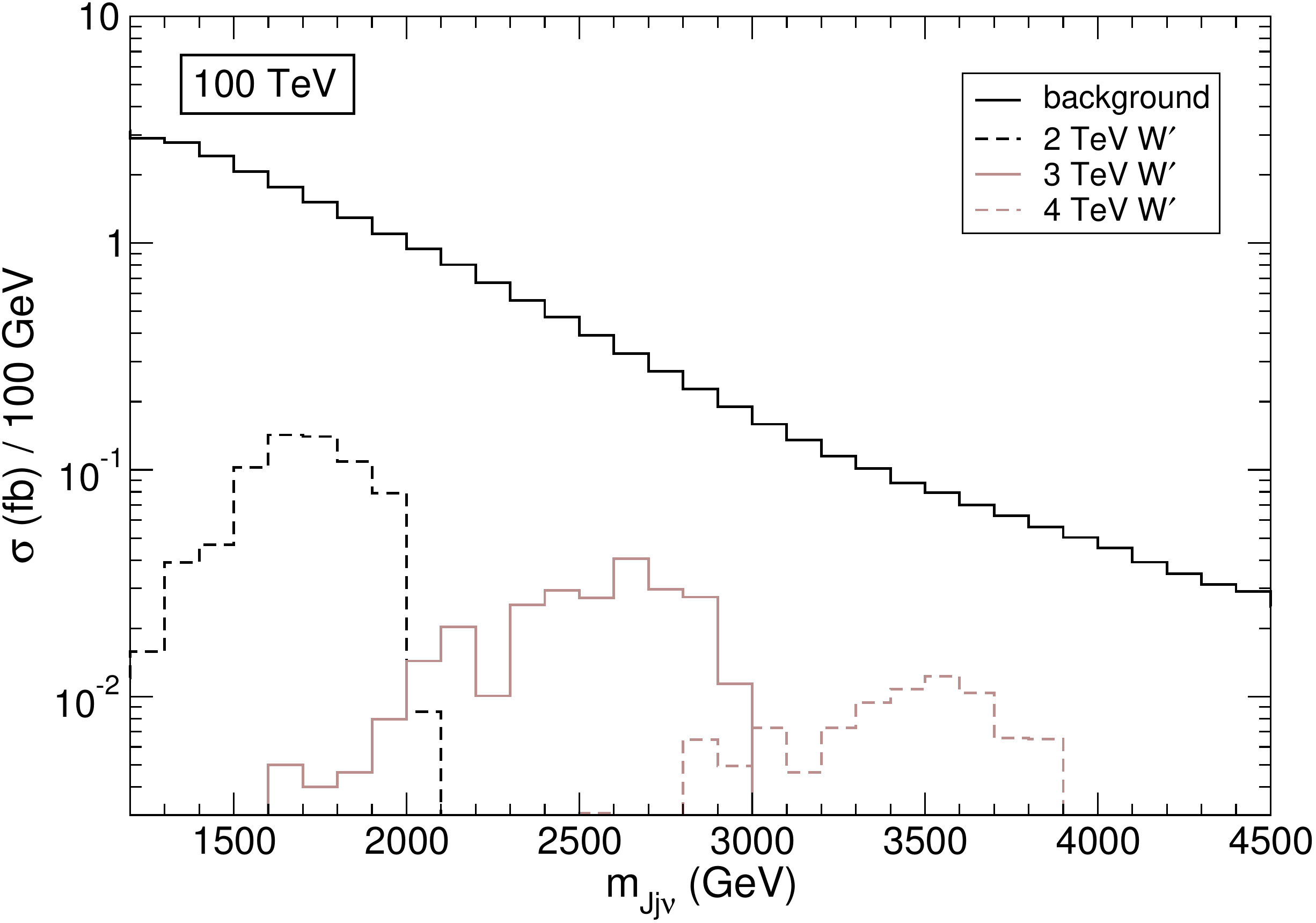}
\end{tabular}
\caption{Signal and background cross sections as a function of the invariant mass $m_{Jj\nu}$ for the $W'$ signals and the SM background, in the $2b$ sample with $z_1 \geq 0.6$.}
\label{fig:dist8}
\end{center}
\end{figure}

We find no improvement in the sensitivity when using the difference of events with positive and negative leptons, therefore we use the full sample with leptons of either sign. A cut on the lepton momentum fraction $z_1 \geq 0.6$ enhances the signal significance, and practically eliminates the dijet and $b \bar b$ backgrounds, see Table~\ref{tab:sig4}.  The $m_{Jj\nu}$ distributions for these signals and the SM background are presented in Fig.~\ref{fig:dist8}.
Notice that the $m_{Jj\nu}$ mass peaks are displaced with respect to the $W'$ mass and the distributions for higher $W'$ mass are very wide. An in-situ jet energy calibration, or a more sophisticated determination of the neutrino longitudinal momentum, would eventually improve the $W'$ mass reconstruction. We have not attempted that, because our conclusions on the observability of the $W'$ signals are not expected to be affected by this calibration of signals and backgrounds. (In the experiment, such calibrations may be performed by comparing the quantities obtained in simulated samples with the original ones.) At high energies the $b$ quark in $W' \to tb$ often radiates an additional jet. We have also tried a $W'$ mass reconstruction taking into account possible additional jets with a separation $\Delta \phi_{Jj} \geq 2.5$ from the top jet (see for example~\cite{delAguila:2009gz}). We find no significant sharpening of the $W'$ reconstructed mass distributions and we do not apply this strategy for simplicity.

In our estimation of the sensitivity of $W'$ searches we do not include systematic uncertainties. The cross section and shape of the dominant $t \bar t$ background can be reliably predicted~\cite{Czakon:2018nun} and the efficiencies for event selection are determined from control regions and sidebands. Current searches in the much more demanding dijet final states, either with the use of jet substructure techniques~\cite{Aad:2019fbh,Sirunyan:2019jbg} or without it~\cite{Aaboud:2017yvp,Sirunyan:2018xlo}, already use control regions to determine directly the background from data. 

\begin{figure}[t]
\begin{center}
\begin{tabular}{c}
\includegraphics[width=8cm]{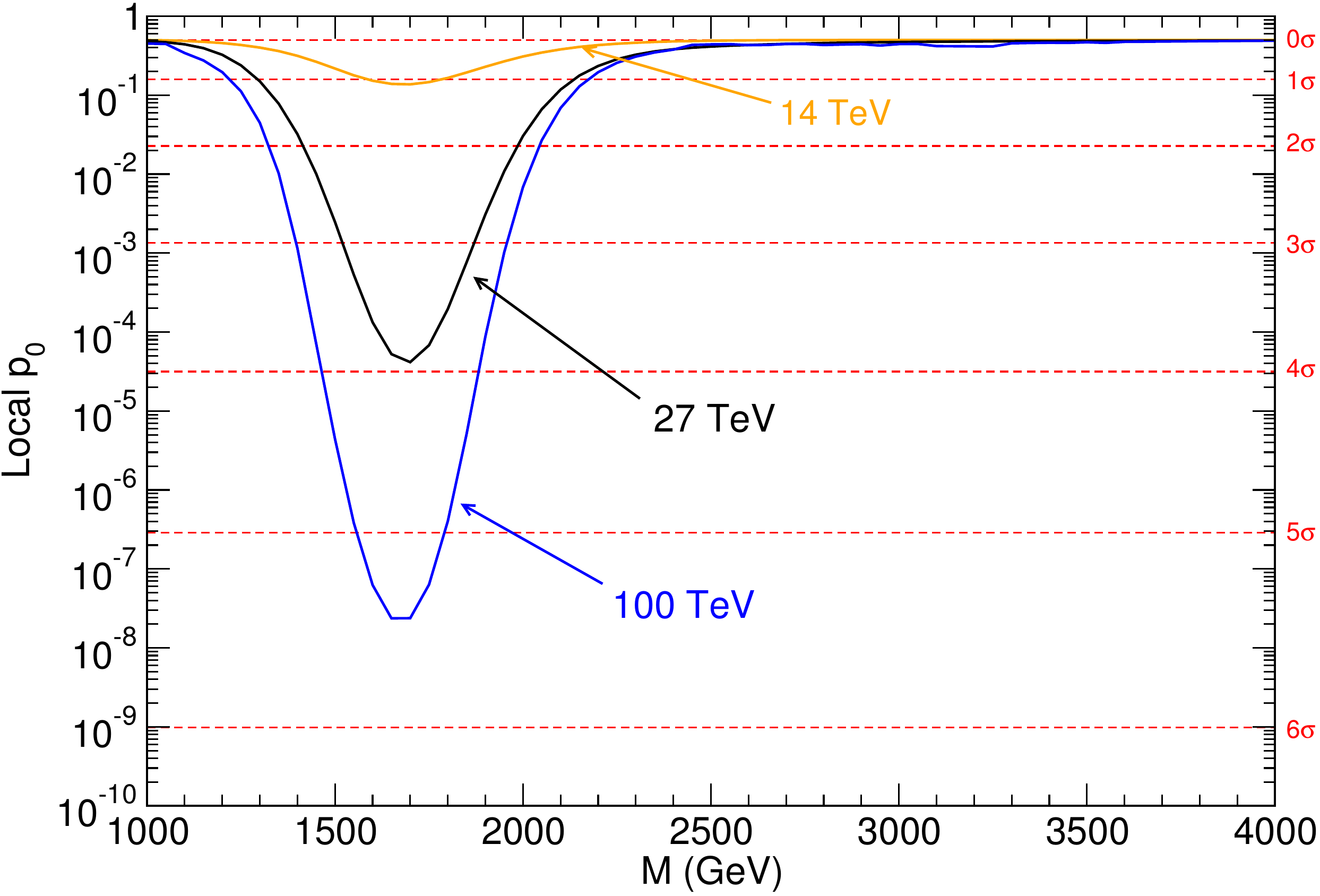} \\
\includegraphics[width=8cm]{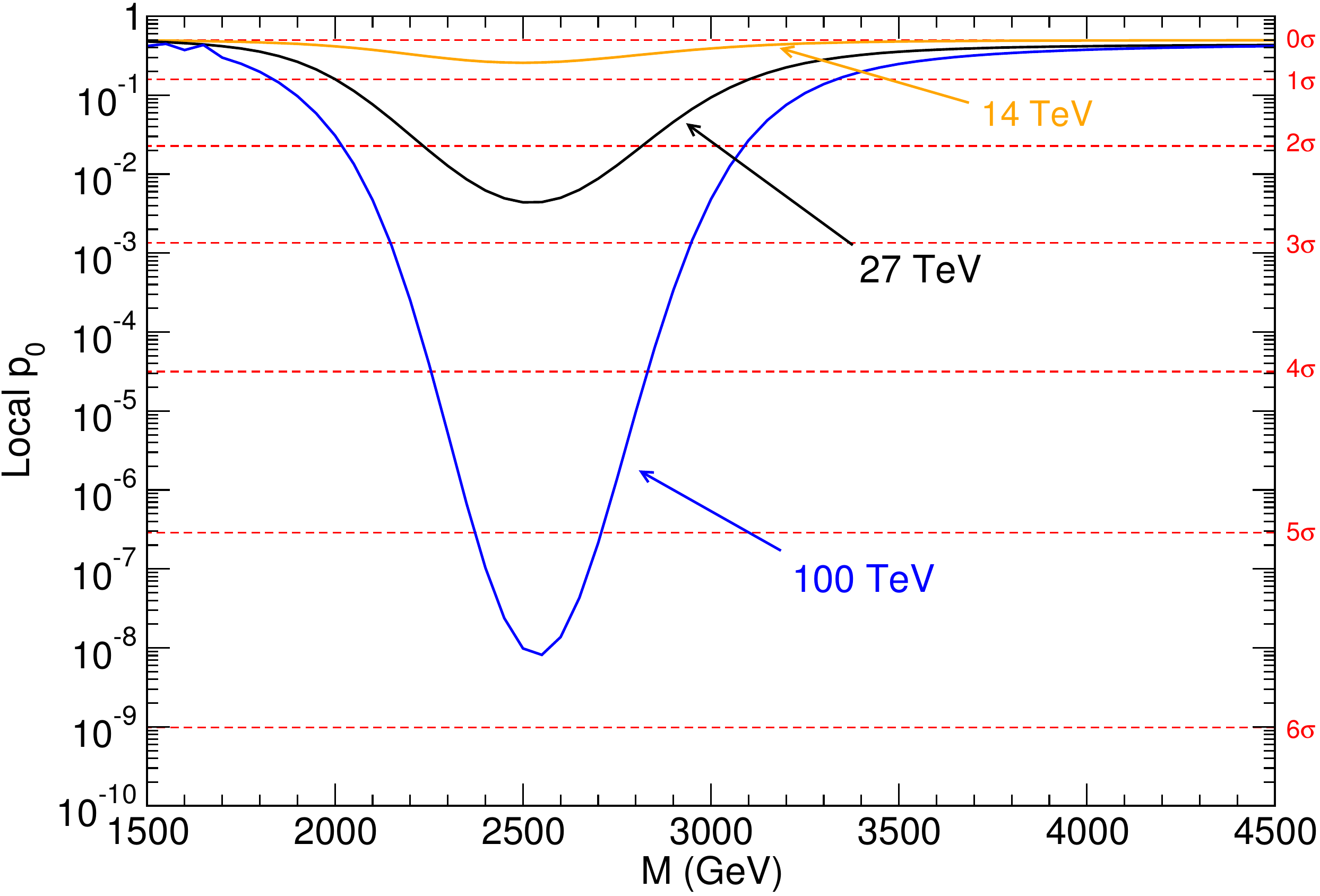} \\
\includegraphics[width=8cm]{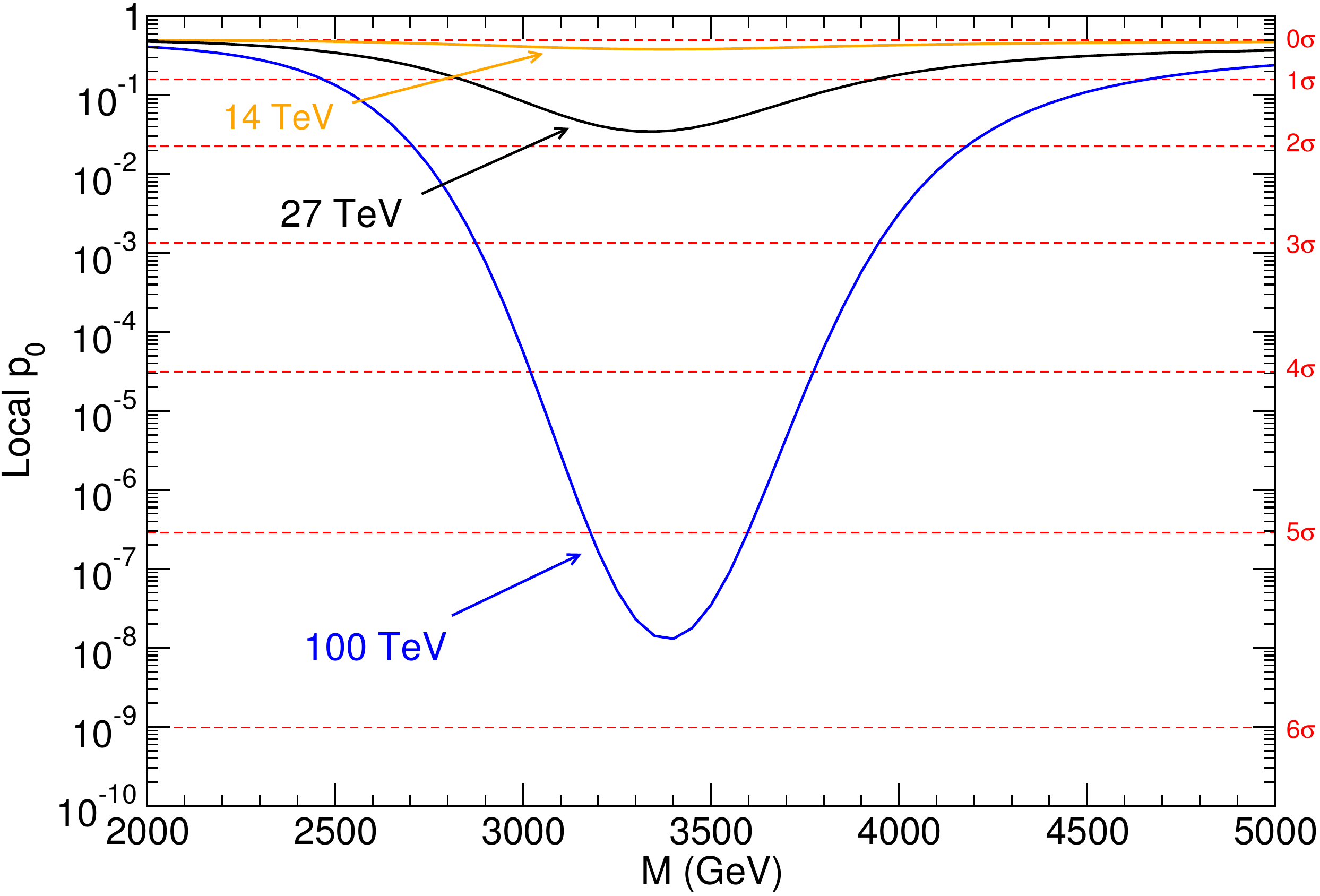}
\end{tabular}
\caption{Local $p$-value of the $W'$ signals for $M_{W'} = 2$ TeV (top), 3 TeV (middle) and 4 TeV (bottom), at the three collider energies considered. The $W'$ couplings for each mass benchmark are given in the text.}
\label{fig:Pval}
\end{center}
\end{figure}

The expected significance of the $W'$ signals are computed by performing likelihood tests for the presence of narrow resonances over the expected background, using the $\text{CL}_\text{s}$ method~\cite{Read:2002hq} with the asymptotic approximation of Ref.~\cite{Cowan:2010js}, and computing the $p$-value corresponding to each hypothesis for the resonance mass. The probability density functions of the potential narrow resonance signals are Gaussians with centre $M$ (i.e. the resonance mass probed) and standard deviation of $0.1 M$. 
The likelihood function is
\begin{equation}
L(\mu) = \prod_{i} \frac{e^{-(b_i+ \mu s_i)} (b_i + \mu s_i) ^{n_i}}{n_i!} \,,
\label{ec:Lik}
\end{equation}
where $i$ runs over the different bins with numbers of observed events $n_i$; $b_i$ is the predicted number of background events and $s_i$ the predicted number of signal events in each bin, and $\mu$ a scale factor.  For each mass hypothesis the value $\mu_b$ that maximises the likelihood function (\ref{ec:Lik}) is calculated, and the local $p$-value is computed as
\begin{equation}
p_0 = 1 - \Phi(\sqrt{2[L(\mu_b)-L(0)]}) \,,
\end{equation}
with
\begin{equation}
\Phi(x) = \frac{1}{2} \left[ 1 + \text{erf}\left(\frac{x}{\sqrt 2}\right) \right] \,.
\end{equation}
The results are presented in Fig.~\ref{fig:Pval} for the three collider energies and: (i) $M_{W'} = 2$ TeV, $g_{W'} = 0.04$ (top);  (ii) $M_{W'} = 3$ TeV, $g_{W'} = 0.05$ (middle); (iii) $M_{W'} = 4$ TeV, $g_{W'} = 0.06$ (bottom). The couplings are chosen so as to have sensitivity near $5\sigma$ at 100 TeV for the three masses studied.

Alternatively, the potential of the three colliders can be compared by calculating the coupling for which there is $5\sigma$ sensitivity  for each $W'$ mass. We summarise these values in Table~\ref{tab:Wplim}. It is seen that, due to the very high statistics, the FCC can probe couplings $2-5$ times smaller than the HL-LHC, despite the smaller $S/B$ ratio visible in Fig.~\ref{fig:dist8}. At high masses the improvement with CM energy is much more pronounced. 
We remark that, for the smaller $W'$ masses, the FCC could gain additional sensitivity by exploring $tb$ final states where the $W'$ itself is boosted at large $p_T$, mimicking the approach used for light-resonance dijet decays in Refs.~\cite{Sirunyan:2017dnz,Aaboud:2018zba}.

\begin{table}[bth]
\begin{center}
\begin{tabular}{lccc}
& 14 TeV & 27 TeV & 100 TeV  \\
$W'$ [2 TeV] & $0.088$ & $0.046$ & 0.038 \\
$W'$ [3 TeV] & $0.15$ & $0.070$ & 0.048 \\
$W'$ [4 TeV] & $0.30$ & $0.11$ & 0.058 
\end{tabular}
\caption{Value of the $W'$ coupling for which there is $5\sigma$ sensitivity, for different masses and collider energies.}
\label{tab:Wplim}
\end{center}
\end{table}

\section{$W'$ mass reach for FCC}
\label{sec:a}

We complement the study of the sensitivity to very weakly-coupled $W'$ bosons, with the determination of the ultimate mass reach for $W'$ bosons with $g_{W'} \sim g$ at the FCC. 

For this study we generate six $t \bar t$ samples in slices of transverse momentum of width 2.5 TeV, starting at $p_T \in [2.5,5]$ TeV and with the last one $p_T \geq 15$ TeV. We ignore the other backgrounds, because as seen in Table~\ref{tab:sig4} they were not relevant in the $2b$ sample with $z_1 \geq 0.6$ and the event selection here is practically the same.
Each background sample has $10^5$ events. We also generate five $W'$ signal samples with $M_{W'}$ ranging from 5 to 30 TeV in steps of 5 TeV, each sample with $6 \times 10^4$ events.

The tagging of multi-TeV $b$ jets based on tracks is problematic, but the performance can be improved by using low-level detector inputs such as hit multiplicity~\cite{btag}. We assume flat $b$ tagging efficiencies of 75\% for $b$ quarks, 20\% for charm and 2\% for light quarks. The precise numbers for charm and light quarks are not crucial for our analysis as the dominant background is $t \bar t$, and the dijet background involving mistags turns out to be very suppressed by the event selection.

We use the same event selection for the $2b$ sample discussed in Section~\ref{sec:2}, except for the jet substructure cut $\tau_{32} \leq 0.7$ which we do not apply because the multi-TeV top jets are very collimated, and to distinguish them from QCD jets a more sophisticated discrimination would be required. We also require $z_1 \geq 0.6$ as done for the analysis in Section~\ref{sec:5}.

\begin{figure}[t]
\begin{center}
\begin{tabular}{c}
\includegraphics[width=8cm]{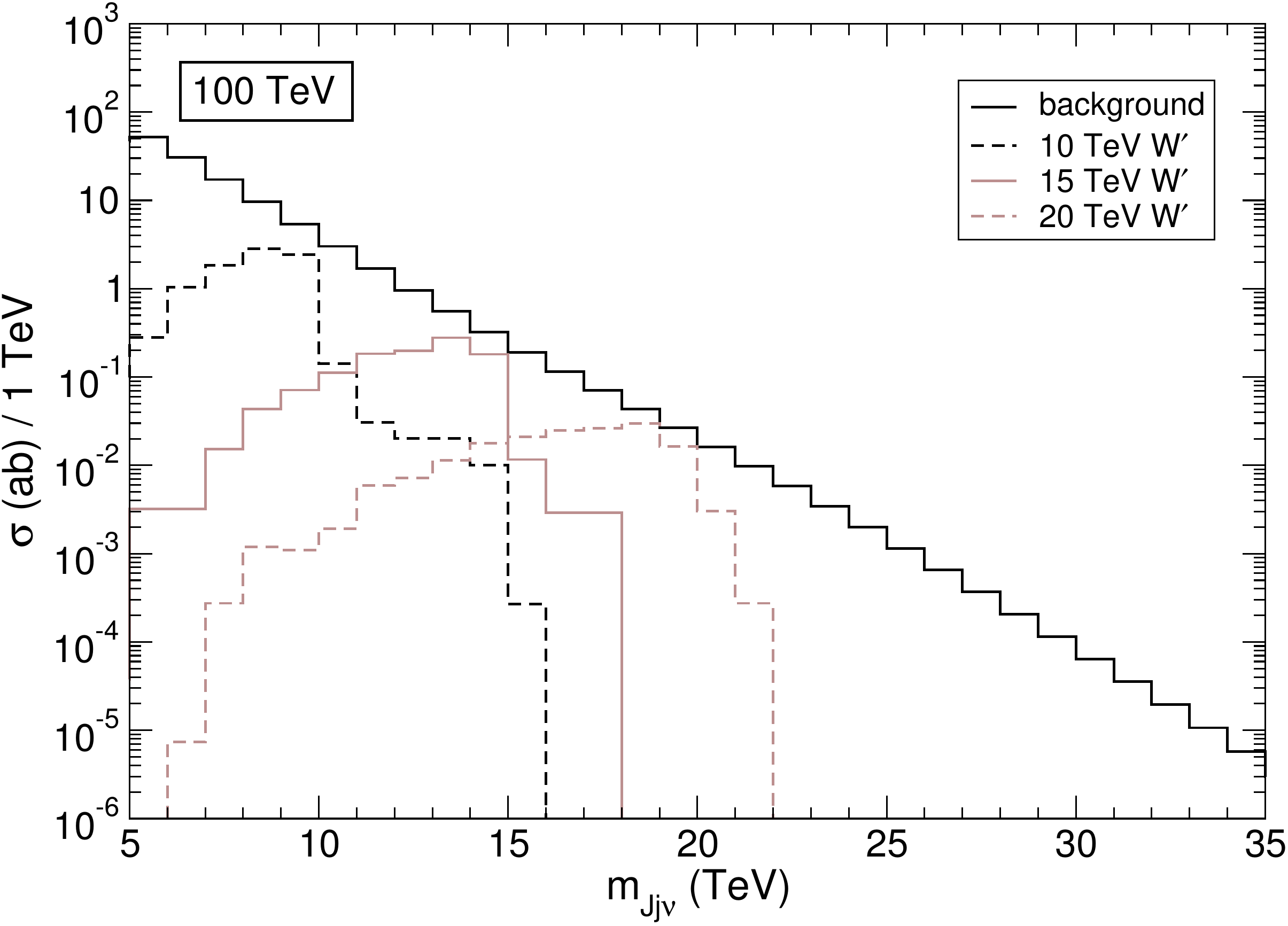}
\end{tabular}
\caption{Signal and background cross sections as a function of the invariant mass $m_{Jj\nu}$ for the $W'$ signals and the SM background, for the high-mass $W'$ search. In the $W'$ signals we set $g_{W'} = 0.1$.}
\label{fig:dist9}
\end{center}
\end{figure}

\begin{figure}[t]
\begin{center}
\begin{tabular}{c}
\includegraphics[width=8cm]{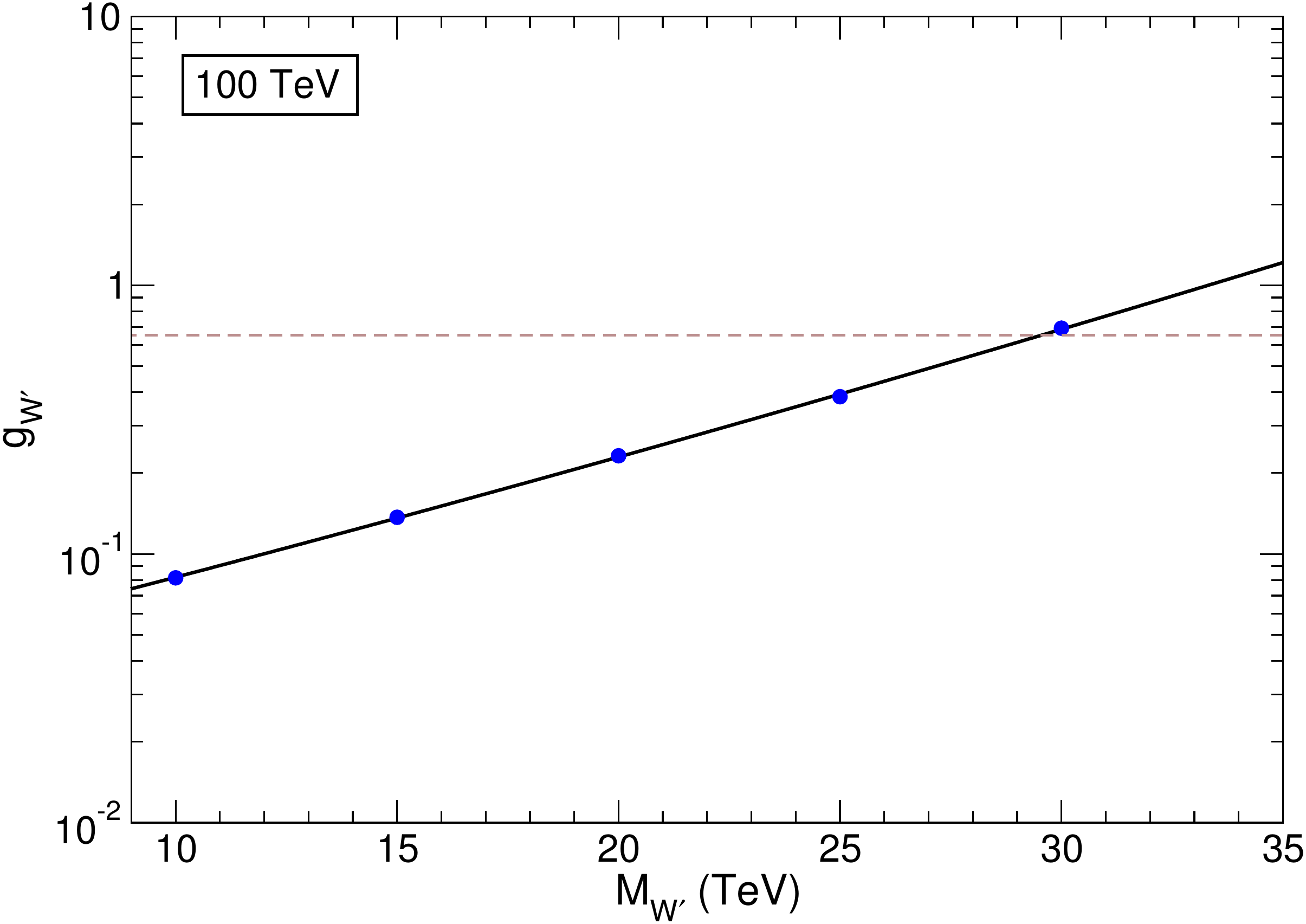}
\end{tabular}
\caption{Coupling $g_{W'}$ for which the sensitivity reaches $5\sigma$, as a function of the $W'$ mass. The blue dots indicate the results from the simulation, and the solid line the fit in Eq.~(\ref{ec:gfit}). The horizontal dashed line corresponds to $g_{W'} = g$.}
\label{fig:g5s}
\end{center}
\end{figure}

The reconstructed $W'$ mass distribution for the background and three $W'$  masses is presented in Fig.~\ref{fig:dist9}. The coupling $g_{W'}$ for which the significance reaches $5\sigma$ is calculated for the five $W'$ masses simulated and shown in Fig.~\ref{fig:g5s}. The values obtained from simulation are well fitted in this mass range with a functional form
\begin{equation}
g_{W'} = A e^{B M+C M^2} \,,
\label{ec:gfit}
\end{equation}
with $A=0.031$, $B= 0.093~\text{TeV}^{-1}$, $C=3.2 \times 10^{-4}~\text{TeV}^{-2}$. From these results, one expects a mass reach of approximately 30 TeV for $g_{W'} = g$.

\section{Discussion}
\label{sec:6}

In Section~\ref{sec:4} we have found that precision measurements of the single top tails can set stringent constraints on anomalous $tbW$ interactions, taking as example the one in Eq.~(\ref{ec:lagr}).
To set this sensivity in context, let us compare the numbers in Eqs.~(\ref{ec:lim14}),(\ref{ec:lim27}) with current limits on the anomalous coupling $g_L$. The limits obtained from measurement of $W$ helicity fractions in $t \bar t$ production are $g_L \in [-0.14,0.11]$~\cite{Aaboud:2016hsq}. Limits from single top production are slightly looser, $|g_L| \lesssim 0.2$~\cite{Khachatryan:2014vma}.  Other angular observables in top decays yield similar constraints, $|g_L| \leq 0.19$~\cite{Aaboud:2017yqf}. At the HL-LHC the precision is expected to improve, assuming a reduction of modeling uncertainties and other systematic uncertainties associated to the Monte Carlo sample size.
The expected limits at 95\% CL are $g_L \in [-0.11,0.08]$ from $W$ helicity fractions and $g_L \in [-0.16,0.19]$ from single top cross sections~\cite{Deliot:2018jts}.\footnote{We thank A. Onofre for providing us with these limits.} 
Therefore, the sensitivity obtained with high-$p_T$ measurements is competitive with precision measurements performed in top production and decays. But, especially, the functional dependence on possible anomalous couplings of the $W$ helicity fractions, inclusive and high-$p_T$ cross sections is quite different, hence the latter are especially interesting when one wants to set global constraints on the $tbW$ vertex including all possible anomalous contributions (four free parameters). This is of special interest given the existence of a flat direction that cannot be probed in the measurements of helicity fractions~\cite{Aguilar-Saavedra:2017wpl}. Because the dependence is different for $t$- and $s$-channel single top production, the two signal regions considered with one and two $b$ tags are also complementary.

The anomalous interaction in (\ref{ec:lagr}) can arise from the dimension-six operator~\cite{AguilarSaavedra:2008zc}
\begin{equation}
O_{dW}^{33} = (\bar q_{L3} \sigma^{\mu \nu} \tau^I b_R) \phi W_{\mu \nu}^I
\label{ec:OdW}
\end{equation}
with $q_{L3} = (t_L \, b_L)^T$, $\phi$ the Higgs doublet, $\tau^I$ the Pauli matrices and $W_{\mu \nu}^I$ the $\text{SU}(2)_L$ field strength tensor. The relation between the effective operator coefficient $C_{dW}^{33}$, the new physics scale $\Lambda$ and the anomalous coupling is
\begin{equation}
g_L = \sqrt 2 C_{dW}^{33*} \frac{v^2}{\Lambda^2} \,,
\end{equation}
with $v=246$ GeV the Higgs vacuum expectation value. Therefore, for $g_L \sim 0.05$ and $C_{dW}^{33} = 1$ the validity of the effective operator approximation requires that the energy scales involved are smaller than $\Lambda = 1.3$ TeV. In order to estimate the energy range that dominates the limits we have repeated the calculations for 14 TeV in the $1b$ sample, setting an upper cut on $p_{T J}$. The 95\% CL limit $|g_L| \leq 0.087$ in Eq.~(\ref{ec:lim14}) changes to 0.088, 0.089, 0.091 and 0.093 when an upper cut $p_{TJ} \leq 1000\,,900\,,800\,,700$ GeV, respectively is set. Therefore, the effective operator approximation is valid even for a small coupling of order unity. For large couplings $C_{dW}^{33} = 16 \pi^2$ the effective operator approximation is valid up to $\Lambda = 16$ TeV.

For completeness, let us also mention that constraints on the coefficient of the operator (\ref{ec:OdW}) from non-top processes can be translated into limits on $g_L$ because the effective operator contains both $tbW$ and $bbZ$ interactions. A global fit to LEP and LHC data in Ref.~\cite{Durieux:2019rbz} obtains a $1\sigma$ sensitivity of $\Delta C_{dW}^{33}/\Lambda^2 \simeq 0.65~\text{TeV}^{-2}$, mainly driven by the measurement of $R_b$ at LEP, which constrains the $bbZ$ term of the effective operator. 
This number can be interpreted as a $1\sigma$ sensitivity of $\Delta g_L = 0.055$, with the caveat that this limit is obtained by measurements of the $bbZ$ interaction, instead of the $tbW$ one as in our case, and the translation is, in principle, only valid within the framework of dimension-six SU(2)-invariant operators.

In Section~\ref{sec:6} we have estimated the sensitivity to weakly-coupled $W'$ bosons in $tb$ final states. We can compare our results with the prospects for the HL-LHC by the ATLAS Collaboration~\cite{ATL-PHYS-PUB-2018-044}. For $M_{W'} = 4$ TeV and $g_{W'} = 0.3$ their sensitivity is nearly $2\sigma$, while our results are more optimistic, reaching $5\sigma$ for these parameters. There are several differences in the analyses, however. Our analysis attempts a more aggressive reduction of the $t \bar t$ background, with a set of cuts that reduces it by a factor of 10, while Ref.~\cite{ATL-PHYS-PUB-2018-044} only sets some loose cuts on the momenta of the jets and the charged lepton. On the other hand, their analysis uses eight different signal regions, splitting the electron and muon samples, by number of $b$ tags and by number of light jets (1 or 2) other than the $b$ jet from the top quark decay. The significances in the individual channels are then combined, which leads to some overall improvement. 
Also, Ref.~\cite{ATL-PHYS-PUB-2018-044} uses next-to-leading order (NLO) cross sections. The $K$ factor (ratio of NLO over LO cross sections) for $W'$ production weakly depends on the $W'$ mass, and for 4 TeV it can be estimated as $K=1.3$~\cite{Sullivan:2002jt}. For $t \bar t$ with $m_{t\bar t} \sim 4$ TeV it is $K=1.4$~\cite{Czakon:2016dgf}. Therefore, the use of NLO cross sections slightly improves the $S/\sqrt{B}$ ratio.

This type of analysis is rare in experimental searches, which usually concentrate on the high-mass end of the spectrum, looking for new resonances with a coupling of order unity. In principle, any search for new resonances is able to spot intermediate mass resonances with small coupling, but in practice the analyses are often optimised for the sensitivity at the high-mass end. With our results in Section~\ref{sec:6} we have demonstrated that, provided the event selection keeps good statistics, future colliders have an excellent potential to explore new $W'$ resonances with coupling as small as a few percent. On the other hand, for couplings of order unity, we showed that the FCC can probe masses up to around 30 TeV.

\begin{acknowledgments}
The work of J.A.A.S. is supported by Spanish Agencia Estatal de Investigaci\'on through the grant `IFT Centro de Excelencia Severo Ochoa SEV-2016-0597' and by MINECO projects FPA 2013-47836-C3-2-P and FPA 2016-78220-C3-1-P (including ERDF).
\end{acknowledgments}


\begin{thebibliography}{99}

\bibitem{Azzi:2019yne}
  P.~Azzi {\it et al.},
  ``Standard Model Physics at the HL-LHC and HE-LHC,''
  arXiv:1902.04070 [hep-ph].

\bibitem{Abada:2019ono}
  A.~Abada {\it et al.} [FCC Collaboration],
  Eur.\ Phys.\ J.\ ST {\bf 228} (2019) no.5,  1109.

  \bibitem{Abada:2019lih}
  A.~Abada {\it et al.} [FCC Collaboration],
  Eur.\ Phys.\ J.\ C {\bf 79} (2019) no.6,  474.

  \bibitem{Benedikt:2018csr}
  A.~Abada {\it et al.} [FCC Collaboration],
  Eur.\ Phys.\ J.\ ST {\bf 228} (2019) no.4,  755.

\bibitem{AguilarSaavedra:2006fy}
  J.~A.~Aguilar-Saavedra, J.~Carvalho, N.~F.~Castro, F.~Veloso and A.~Onofre,
  Eur.\ Phys.\ J.\ C {\bf 50} (2007) 519
  [hep-ph/0605190].

\bibitem{AguilarSaavedra:2008gt}
  J.~A.~Aguilar-Saavedra,
  Nucl.\ Phys.\ B {\bf 804} (2008) 160
  [arXiv:0803.3810 [hep-ph]].

\bibitem{Godbole:2018wfy}
  R.~M.~Godbole, M.~E.~Peskin, S.~D.~Rindani and R.~K.~Singh,
  Phys.\ Lett.\ B {\bf 790} (2019) 322
  [arXiv:1809.06285 [hep-ph]].

\bibitem{Alwall:2014hca}
 J.~Alwall {\it et al.},
 JHEP {\bf 1407} (2014) 079
 [arXiv:1405.0301 [hep-ph]].

\bibitem{Sjostrand:2007gs}
  T.~Sjostrand, S.~Mrenna and P.~Z.~Skands,
  Comput.\ Phys.\ Commun.\  {\bf 178} (2008) 852
  [arXiv:0710.3820 [hep-ph]].

\bibitem{deFavereau:2013fsa}
  J.~de Favereau {\it et al.} [DELPHES 3 Collaboration],
  JHEP {\bf 1402} (2014) 057
  [arXiv:1307.6346 [hep-ex]].

\bibitem{card}
See https://twiki.cern.ch/twiki/bin/view/LHCPhysics/ HLHEWG\_MC


\bibitem{Cacciari:2008gp}
  M.~Cacciari, G.~P.~Salam and G.~Soyez,
  JHEP {\bf 04} (2008) 063
  [arXiv:0802.1189 [hep-ph]].

\bibitem{Krohn:2009th}
  D.~Krohn, J.~Thaler and L.~T.~Wang,
  JHEP {\bf 1002} (2010) 084
  [arXiv:0912.1342 [hep-ph]].

\bibitem{Larkoski:2014wba}
  A.~J.~Larkoski, S.~Marzani, G.~Soyez and J.~Thaler,
  JHEP {\bf 1405} (2014) 146
  [arXiv:1402.2657 [hep-ph]].

\bibitem{Cacciari:2011ma}
  M.~Cacciari, G.~P.~Salam and G.~Soyez,
  Eur.\ Phys.\ J.\ C {\bf 72} (2012) 1896
  [arXiv:1111.6097 [hep-ph]].

\bibitem{Aguilar-Saavedra:2017vka}
  J.~A.~Aguilar-Saavedra,
  Eur.\ Phys.\ J.\ C {\bf 77} (2017) no.11,  769
  [arXiv:1709.03975 [hep-ph]].

\bibitem{Bertolini:2014bba}
  D.~Bertolini, P.~Harris, M.~Low and N.~Tran,
  JHEP {\bf 1410} (2014) 059
  [arXiv:1407.6013 [hep-ph]].

\bibitem{Cacciari:2014gra}
  M.~Cacciari, G.~P.~Salam and G.~Soyez,
  Eur.\ Phys.\ J.\ C {\bf 75} (2015) no.2,  59
  [arXiv:1407.0408 [hep-ph]].

\bibitem{Berta:2014eza}
  P.~Berta, M.~Spousta, D.~W.~Miller and R.~Leitner,
  JHEP {\bf 1406} (2014) 092
  [arXiv:1403.3108 [hep-ex]].

\bibitem{Aaboud:2017cxo}
  M.~Aaboud {\it et al.} [ATLAS Collaboration],
  JHEP {\bf 1803} (2018) 174
  [arXiv:1712.06518 [hep-ex]].

\bibitem{Thaler:2010tr}
  J.~Thaler and K.~Van Tilburg,
  JHEP {\bf 1103} (2011) 015
  [arXiv:1011.2268 [hep-ph]].

\bibitem{Komiske:2017aww}
  P.~T.~Komiske, E.~M.~Metodiev and J.~Thaler,
  JHEP {\bf 1804} (2018) 013
  [arXiv:1712.07124 [hep-ph]].

\bibitem{Aguilar-Saavedra:2014iga} 
  J.~A.~Aguilar-Saavedra, B.~Fuks and M.~L.~Mangano,
  Phys.\ Rev.\ D {\bf 91}, 094021 (2015)
  [arXiv:1412.6654 [hep-ph]].

\bibitem{Sirunyan:2018rlu} 
  A.~M.~Sirunyan {\it et al.} [CMS Collaboration],
  arXiv:1812.10514 [hep-ex].





\bibitem{AguilarSaavedra:2010nx}
  J.~A.~Aguilar-Saavedra and J.~Bernabeu,
  Nucl.\ Phys.\ B {\bf 840} (2010) 349
  [arXiv:1005.5382 [hep-ph]].

\bibitem{Czakon:2018nun} 
  M.~Czakon, A.~Ferroglia, D.~Heymes, A.~Mitov, B.~D.~Pecjak, D.~J.~Scott, X.~Wang and L.~L.~Yang,
  JHEP {\bf 1805}, 149 (2018)
  [arXiv:1803.07623 [hep-ph]].
  
  

\bibitem{Aad:2019fbh} 
  G.~Aad {\it et al.} [ATLAS Collaboration],
  arXiv:1906.08589 [hep-ex].


\bibitem{Sirunyan:2019jbg} 
  A.~M.~Sirunyan {\it et al.} [CMS Collaboration],
  arXiv:1906.05977 [hep-ex].


\bibitem{Aaboud:2017yvp}
  M.~Aaboud {\it et al.} [ATLAS Collaboration],
  Phys.\ Rev.\ D {\bf 96} (2017) no.5,  052004
  [arXiv:1703.09127 [hep-ex]].

\bibitem{Sirunyan:2018xlo} 
  A.~M.~Sirunyan {\it et al.} [CMS Collaboration],
  JHEP {\bf 1808}, 130 (2018)
  [arXiv:1806.00843 [hep-ex]].
  

\bibitem{delAguila:2010mx}
  F.~del Aguila, J.~de Blas and M.~P\'erez-Victoria,
  JHEP {\bf 1009} (2010) 033
  [arXiv:1005.3998 [hep-ph]].

\bibitem{Sirunyan:2017vkm}
  A.~M.~Sirunyan {\it et al.} [CMS Collaboration],
  Phys.\ Lett.\ B {\bf 777} (2018) 39
  [arXiv:1708.08539 [hep-ex]].

\bibitem{Aaboud:2018jux}
  M.~Aaboud {\it et al.} [ATLAS Collaboration],
  Phys.\ Lett.\ B {\bf 788} (2019) 347
  [arXiv:1807.10473 [hep-ex]].
  
\bibitem{delAguila:2009gz} 
  F.~del Aguila, J.~A.~Aguilar-Saavedra, M.~Moretti, F.~Piccinini, R.~Pittau and M.~Treccani,
  Phys.\ Lett.\ B {\bf 685}, 302 (2010)
  [arXiv:0912.3799 [hep-ph]].

  
  
\bibitem{Read:2002hq}
  A.~L.~Read,
  J.\ Phys.\ G {\bf 28} (2002) 2693.

\bibitem{Cowan:2010js}
  G.~Cowan, K.~Cranmer, E.~Gross and O.~Vitells,
  Eur.\ Phys.\ J.\ C {\bf 71} (2011) 1554
   Erratum: Eur.\ Phys.\ J.\ C {\bf 73} (2013) 2501
  [arXiv:1007.1727 [physics.data-an]].


  
  





\bibitem{Aaboud:2016hsq}
  M.~Aaboud {\it et al.} [ATLAS Collaboration],
  Eur.\ Phys.\ J.\ C {\bf 77} (2017) no.4,  264
  [arXiv:1612.02577 [hep-ex]].

\bibitem{Khachatryan:2014vma}
  V.~Khachatryan {\it et al.} [CMS Collaboration],
  JHEP {\bf 1501} (2015) 053
  [arXiv:1410.1154 [hep-ex]].

\bibitem{Aaboud:2017yqf}
  M.~Aaboud {\it et al.} [ATLAS Collaboration],
  JHEP {\bf 1712} (2017) 017
  [arXiv:1707.05393 [hep-ex]].


\bibitem{Deliot:2018jts} 
  F.~D\'eliot, M.~C.~N.~Fiolhais and A.~Onofre,
  Mod.\ Phys.\ Lett.\ A {\bf 34}, no. 18, 1950142 (2019)
  [arXiv:1811.02492 [hep-ph]].

\bibitem{Durieux:2019rbz} 
  G.~Durieux, A.~Irles, V.~Miralles, A.~Peñuelas, R.~P\"oschl, M.~Perell\'o and M.~Vos,
  arXiv:1907.10619 [hep-ph].


\bibitem{AguilarSaavedra:2008zc}
  J.~A.~Aguilar-Saavedra,
  Nucl.\ Phys.\ B {\bf 812} (2009) 181
  [arXiv:0811.3842 [hep-ph]].

\bibitem{Aguilar-Saavedra:2017wpl}
  J.~A.~Aguilar-Saavedra, J.~Boudreau, C.~Escobar and J.~Mueller,
  Eur.\ Phys.\ J.\ C {\bf 77} (2017) no.3,  200
  [arXiv:1702.03297 [hep-ph]].

\bibitem{ATL-PHYS-PUB-2018-044}
ATLAS Collaboration, note ATL-PHYS-PUB-2018-044


\bibitem{Sullivan:2002jt} 
  Z.~Sullivan,
  Phys.\ Rev.\ D {\bf 66}, 075011 (2002)
  [hep-ph/0207290].




\bibitem{Czakon:2016dgf} 
  M.~Czakon, D.~Heymes and A.~Mitov,
  JHEP {\bf 1704}, 071 (2017)
  [arXiv:1606.03350 [hep-ph]].


\bibitem{Sirunyan:2017dnz}
  A.~M.~Sirunyan {\it et al.} [CMS Collaboration],
  Phys.\ Rev.\ Lett.\  {\bf 119} (2017) no.11,  111802
  [arXiv:1705.10532 [hep-ex]].
\bibitem{Aaboud:2018zba}
  M.~Aaboud {\it et al.} [ATLAS Collaboration],
  Phys.\ Lett.\ B {\bf 788} (2019) 316
  [arXiv:1801.08769 [hep-ex]].
  
  \bibitem{btag}
E. P\'erez Codina and P. G. Roloff,
Tech. Rep. CERN-ACC-2018-0023, https://cds.cern.ch/record/2631478.


\end{thebibliography}
\end{document}